\documentclass[aps,prd,reprint,superscriptaddress,twocolumn,preprintnumbers,floatfix,nofootinbib,amsmath,amssymb]{revtex4-1}

\usepackage{graphicx}
\usepackage{epstopdf}
\usepackage{mathrsfs}
\usepackage{amssymb}
\usepackage{verbatim}
\usepackage{color}
\usepackage[dvipsnames]{xcolor}
\usepackage{multirow}
\usepackage{amsmath}
\usepackage{subfig}
 \usepackage{slashed}
\usepackage{float}
\usepackage[normalem]{ulem}
\usepackage{sidecap}
\usepackage{mathtools}

\usepackage{hyperref}
\hypersetup{pdfstartview=FitV,colorlinks=true,linkcolor=blue,citecolor=red,filecolor=black,urlcolor=blue}

\usepackage{diagbox}

\usepackage[T1]{fontenc}
\usepackage[utf8]{inputenc}

\newcommand{\gev}{\, {\rm GeV}}
\newcommand{\beq}{\begin{equation}}
\newcommand{\eeq}{\end{equation}}
\newcommand{\bea}{\begin{eqnarray}}
\newcommand{\eea}{\end{eqnarray}}

\newcommand{\ttiny}[1]{\text{\tiny{$#1$}}}

\newcommand{\T}{\mathcal{T}}

\def\dm{\text{\tiny{DM}}}
\def\sm{\text{\tiny{SM}}}

\def\rh{\text{\tiny{RH}}}
\def\max{\text{\tiny{MAX}}}

\def\dmS{S} 
\def\dmF{\chi} 
\def\dmV{\mathcal{V}} 
\def\dm{\ttiny{DM}}
\def\sm{\ttiny{SM}}

\def\r{t}
\def\i{a}
\def\mre{m_\r}
\def\mim{m_\i}
\def\Gre{\G_\r}
\def\Gim{\G_\i}
\def\a{\alpha}
\def\b{\beta}
\def\g{\gamma}
\def\l{\lambda}
\def\m{\mu}

\def\G{\Gamma}
\def\L{\Lambda}
\def\O{\Omega}

\def\to{\rightarrow}

\def\M{\mathcal{M}}

\def\Z{\mathcal{Z}}

\usepackage[parfill]{parskip}

\begin{document}\sloppy 

\vspace*{1mm}

\title{Moduli Portal Dark Matter}
\author{Debtosh Chowdhury}
\email{debtosh.chowdhury@polytechnique.edu}
\affiliation{Centre de Physique Th{\'e}orique, {\'E}cole Polytechnique, CNRS, Université Paris-Saclay, 91128 Palaiseau Cedex, France}
\affiliation{Laboratoire de Physique Th{\'e}orique (UMR8627), CNRS, Univ.~Paris-Sud, Universit{\'e} Paris-Saclay, 91405 Orsay, France}

\author{Emilian Dudas}
\email{emilian.dudas@polytechnique.edu}
\affiliation{Centre de Physique Th{\'e}orique, {\'E}cole Polytechnique, CNRS, Université Paris-Saclay, 91128 Palaiseau Cedex, France}

\author{Maíra Dutra}
\email{maira.dutra@th.u-psud.fr}
\affiliation{Laboratoire de Physique Th{\'e}orique (UMR8627), CNRS, Univ.~Paris-Sud, Universit{\'e} Paris-Saclay, 91405 Orsay, France}

\author{Yann Mambrini}
\email{yann.mambrini@th.u-psud.fr}
\affiliation{Laboratoire de Physique Th{\'e}orique (UMR8627), CNRS, Univ.~Paris-Sud, Universit{\'e} Paris-Saclay, 91405 Orsay, France}


\begin{abstract} 
We show that moduli fields as mediators between the Standard Model and the dark sector can naturally lead to the observed relic abundance. Indeed, even if moduli are very massive, the nature of their couplings with matter and gauge fields allows producing a sufficiently large amount of dark matter in the early Universe through the freeze-in mechanism. Moreover, the complex nature of the moduli fields whose real and imaginary part couple differently to the thermal bath gives an interesting and unusual phenomenology compared to other freeze-in models of that type.
\end{abstract}

\date{\today}
\preprint{LPT--Orsay 18-83}
\preprint{CPHT-RR099.102018}

\maketitle

\setcounter{equation}{0}

\section{Introduction}
\label{secI}

Despite indirect but clear evidences \cite{planck_collaboration_planck_2018} of the presence of large amount of dark matter in our Universe, its nature still remains elusive. The absence of any signal in direct detection experiments like XENON \cite{xenon100_collaboration_dark_2012_ok}, LUX \cite {akerib_results_2017_ok}, and PANDAX\cite{fu_spin-dependent_2017_ok} questions the weakly coupled dark matter paradigm. Simplest extensions as Higgs-portal  \cite{casas_reopening_2017,*djouadi_implications_2012,*djouadi_direct_2013,*lebedev_vector_2012,*mambrini_higgs_2011}, $Z$-portal \cite{ellis_statistical_2018,*arcadi_z-portal_2015,*kearney_$z$_2017,*escudero_toward_2016}, or even $Z'$-portal \cite{alves_dark_2014,*lebedev_axial_2014,*arcadi_invisible_2014} have large part of their parameter space (if not all) excluded when combining direct, indirect and accelerator searches (for a review on WIMP searches and models, see \cite{arcadi_waning_2018}). This uncomfortable situation justifies the need to look for different scenarios, allowing feeble couplings, or the possibility of dark matter production at the very early stages of reheating. Such mechanisms are usually dubbed under the acronym FIMP \cite{hall_freeze-production_2010,*chu_four_2012,*chu_thermal_2014} for feebly interacting massive particles (see \cite{bernal_dawn_2017} for a review).

In this context, several models have been studied
and it has been confirmed that dark matter production is naturally feasible in different setup like SO(10) unified
construction \cite{mambrini_gauge_2013,*mambrini_dark_2015,*mambrini_vacuum_2016}, U$(1)'$ anomaly free models \cite{bhattacharyya_freezing-dark_2018}, spin-2 portal \cite{bernal_spin-2_2018}
or high scale supersymmetry (SUSY) \cite{benakli_minimal_2017,*dudas_case_2017,*dudas_inflation_2017,*dudas_gravitino_2018}. In all these models, it has been shown that effects of non-instantaneous reheating \cite{garcia_enhancement_2017} and non-instantaneous thermalization \cite{garcia_pre-thermalization_2018} should be considered with care.
It seemed then interesting to study constructions with
massive scalar moduli fields abundantly present in supergravity (SUGRA) and string theory extensions, and to check if they can play the role of a mediator between the dark sector and the Standard Model. Indeed moduli fields couple generically to Standard Model fields through higher-dimensional operators, mostly of derivative interactions type. As a consequence, mechanisms  implying moduli are important at high energies/temperatures, being therefore potentially relevant for the freeze-in mechanism.

Moduli fields appear in many extensions of the Standard Model. Indeed, in any higher-dimensional supergravity or string theory extension of the Standard Model, there are scalar fields coming from the compactification of the higher-dimensional metric, dilaton or various antisymmetric tensors. In particular, internal volumes and shapes and their axionic partners are abundant in such constructions. Most of them are flat directions at tree-level and get potentials and therefore masses by various perturbative and non-perturbative effects. Their resulting masses and vacuum expectation values  are model dependent and will be taken as free parameters in what follows. Their vacuum expectation determine values of four-dimensional parameters: gauge and Yukawa couplings, wave functions of various fields and Planck mass. If one assumes that the low-energy theory, obtained after their decoupling, is the Standard Model or a phenomenologically motivated extension of it, then their couplings can be obtained by starting from the low-energy theory and expanding the low-energy parameters in a power series. If moduli fields are heavier than the reheating temperature, then they can be safely replaced by their vacuum expectation values (vevs) and the low-energy theory is just the Standard Model or appropriate extension. If they are lighter however, they can lead to various physical effects. This strategy was used in early papers \cite{taylor_dilaton_1988,damour_string_1994,binetruy_nambu_1994,*Binetruy:1994bn,*Binetruy:1995nt, kounnas_towards_1994,dimopoulos_macroscopic_1996,antoniadis_millimetre-range_1997,Feruglio:2017spp,*Criado:2018thu} in order to study various low-energy effects of the moduli fields. Recently there have been works in the direction where moduli were considered as dark matter candidate \cite{Kusenko:2012ch} or production of dark matter from moduli decay in the early Universe \cite{Acharya:2008bk,Acharya:2009zt,Moroi:1999zb,Allahverdi:2013tca}.
The present work is intended to complete the list of interesting consequences of moduli fields by studying their possible role as mediators between the Standard Model and dark matter sector.

The paper is organized as follows. In section \ref{secII},  we describe the model under consideration. In section \ref{secIII}, we discuss bounds on the moduli masses coming from cosmology. In section \ref{secIV}, we outline the dark matter relic abundance through the freeze-in mechanism in the early Universe, taking into account non-instantaneous reheating. Section \ref{secV} is devoted to  the computation of the the dark matter production rate in the early Universe in our model and we delineate the parameter space for the model in consideration. In section \ref{secVI}, we summarize the main results of our work and conclude by highlighting the new aspects that emerged from our analysis.

\section{The model}
\label{secII}

Let $\L$ being the new physics scale (that can be string scale, unification scale or SUSY/SUGRA breaking scale, for instance). Consistency of the effective field theory requires that $\L$ is the largest mass scale of theory, in particular larger than dark matter or mediator masses, and the maximum temperature after reheating. One can then define the couplings of the complex modulus field\footnote{We will consider only one modulus field throughout our work. Generalization to several fields is straightforward.} $\T$ decomposed as $\T \equiv \r + i \i$, to the Standard Model field $k$ by expanding the wave-functions $\mathcal{Z}_k$:
\begin{equation}\label{Eq:1}
\mathcal{Z}_k(\T, \bar{\T})  \approx 1 + \frac{c_k}{\L} \T + \frac{d_k}{\L} \bar{\T} \equiv 1 + \frac{\a_k}{\L} \r + i \frac{\b_k}{\L} \i \, ,
\end{equation}
where $c_k$ and $d_k$ are real coefficients of order one and we defined the couplings to the real and imaginary components of $\T$ as $\a_k = c_k+d_k$ and $\b_k = c_k-d_k$ respectively\footnote{For simplicity, throughout our work, we will consider CP-conserving Lagrangians and therefore real coefficients in the couplings of moduli to matter. Extension to CP-violating couplings is interesting but beyond the goal of our paper.}. We can then express generic couplings of the moduli fields to the Standard Model sector as
\begin{align}\label{Eq:2}
{\cal L}_\T^{SM} \supset & ~\mathcal{Z}_H |D_\mu H|^2 - \mu^2(\T,\bar \T)|H|^2 - \lambda(\T,\bar \T) |H|^4 \notag \\ 
& + \frac{1}{2} \left(\mathcal{Z}_L \bar f_L i 
\slashed{D} f_L 
+ \mathcal{Z}_R \bar f_R i 
\slashed{D} f_R + \text{h.c.}\right) \notag  \\
& -\frac{1}{4}\Z_G~ G_{\mu \nu} G^{\mu \nu}
- \Z^\prime_{G} ~G_{\mu \nu} \tilde G^{\mu \nu} \,,  
\end{align}
where $\Z_H (= 1 + \frac{\a_H}{\L}\r$), $\Z_{L,R} (= 1 + \frac{\a_{L,R}}{\L} \r + i \frac{\b_{L,R}}{\L} \i$), $\Z_G (= 1 + \frac{\a_G}{\L} \r$) and $\Z^\prime_{G} (= \frac{\b_G}{\L} \i)$ are the wave functions of the scalar ($H$), fermionic ($f$) and (Abelian and non-Abelian) gauge ($G_\mu$) fields of the Standard Model respectively. In the above equation, $G_{\mu \nu}$ is the field strength tensor of the gauge field ($G_\mu$) and $\tilde{G}^{\mu \nu} (= \frac{1}{2} \epsilon^{\mu \nu \rho \sigma} G_{\rho \sigma}$) is its dual field strength tensor.

From the first line of Eq.\eqref{Eq:2} we see that the scalar potential depends on the mass parameter $\mu$ which is also a function of the moduli fields. Parametrizing the contribution of the moduli to the $\mu$-parameter in a similar fashion as in Eq.\eqref{Eq:1} we can write,
\begin{align}\label{caseA}
    \mu^2 = \mu_0^2 \left( 1 + \frac{\a_H}{\Lambda}~ t \right) \, ,
\end{align}
with $\mu_0$ being the SM $\mu$-parameter that reproduces the observed Higgs mass at the electroweak scale. As $\L$ is the highest scale in the theory, contribution to the Higgs mass due to the moduli is small. On the other hand there is a second possibility that the $\mu$-parameter gets generated at a scale ($\sqrt{\langle F \rangle}$) close to the Planck scale. In this case the effective $\mu$-parameter can be written  
\beq\label{caseB}
\mu^2 = \mu_0^2 + \frac{\langle F \rangle}{M_P}\r \, ,
\eeq
where $\langle F \rangle$ is the vev of the ``spurion'' field. In this case, one needs a considerable amount of cancellation or fine-tuning between the two contributions in Eq.\eqref{caseB} to reproduce the observed Higgs mass. In contrast to the Eq.\eqref{caseA}, in this case the coupling of $
\r$ to the Higgs is quite large. This leads to the fact that the width of $\r$ could be larger than the mass of $\r$ unless we demand that the width to be at most the mass of $\r$. This sets an upper bound on the spurion vev, $\langle F \rangle \lesssim m_\r M_P$, where $m_\r$ is the mass of $\r$. Throughout our analysis, we will consider the case of Eq.\eqref{caseA} unless otherwise stated.



The effective interactions between the components of the moduli and SM fields, at the first order in $1/\L$, reads
\begin{align}\label{Eq:leff}
{\cal L}_\T^{SM} \supset 
& \frac{\alpha_H}{\Lambda}~\r~|D_\mu H|^2 - \frac{\a_H}{\Lambda}~\m_0^2~\r~|H|^2 \notag \\
& + \left(\frac{1}{2\Lambda} ~\r~ \bar f i \gamma^\mu (\a_\text{V}^f - \a_\text{A}^f \g_5) D_\mu f + \text{h.c.}\right) \notag \\
& + \frac{1}{2\Lambda} \partial_\mu \i ~\bar f \gamma^\mu (\b_\text{V}^f - \b_\text{A}^f \g_5) f  \\
& - \frac{1}{4}\frac{\alpha_G}{\Lambda}~\r~G_{\mu \nu} G^{\mu \nu} + 2 \frac{\beta_G}{\Lambda} ~\partial_\mu \i~\epsilon^{\mu \nu \rho \sigma} G_\nu \partial_\rho G_\sigma \, , \notag
\end{align}
where we have identified the chiral couplings as $\a_\text{V}^f = (\a_L+\a_R)/2$ and $\a_\text{A}^f = (\a_L-\a_R)/2$, with analogous definitions for the couplings of the imaginary part of the moduli. 

Before we proceed further we can make some remarks after having a quick look at Eq.\eqref{Eq:leff}:
\begin{itemize}
\item Since the kinetic term of Higgs needs to be real, the Higgs sector only couples with the real part of the modulus field $\r$.
\item One observes that the Lagrangian in Eq.\eqref{Eq:leff} is invariant under a shift in the imaginary part of the moduli ($\i \rightarrow \i$ + const.). This can also be observed in SUGRA models for instance, where the Kähler metric depends explicitly on the combination $\T + \bar \T$. In other words the Lagrangian could be written by imposing the shift symmetry from the beginning\footnote{ This argument is not valid if one includes Yukawa-like couplings in the action.  However their contribution to dark matter production in the early Universe is negligible.}. As a consequence, the nature of the couplings of $\i$ to the Standard Model fields differs from the couplings of $\r$ ($\i$ develops naturally derivative-type couplings). 
\item We kept in Eq.\eqref{Eq:leff} only the leading operators which contributes  to 3-point functions  (so the partial derivative). This is because $2 \rightarrow 2$ production of dark matter will dominate on the $3 \rightarrow 2$ production\footnote{$3 \rightarrow 2$ processes can dominate if  $2 \rightarrow 2$ channels are suppressed, and when strong interactions are generated to compensate the loss induced by the reduced phase space \cite{Choi:2017zww}.}. 
\end{itemize}

By analogy, one can write the same type of couplings in the dark sector. In our study, we will distinguish three cases of dark mater: scalar ($\dmS$), fermionic ($\dmF$) and vectorial ($\dmV$). Their interactions with moduli reads
\begin{equation}\label{Eq:dms}
\mathcal{L}_{\T}^{\dmS} =  \frac{\alpha_\dmS}{\Lambda}~\r~|\partial_\mu \dmS|^2 \, ,
\end{equation}
\begin{equation}\begin{split}\label{Eq:dmf}
\mathcal{L}_{\T}^{\dmF} =\ & ~ \left(\frac{1}{2\L} ~\r~ \bar \dmF i \gamma^\mu (\a_\text{V}^\dmF - \a_\text{A}^\dmF \gamma_5) ~\partial_\mu \dmF + \text{h.c.} \right) \\ 
& + \frac{1}{2\L} ~\partial_\mu \i ~\bar \dmF   \gamma^\mu (\b_\text{V}^\dmF - \b_\text{A}^\dmF \gamma_5) \dmF \, ,
\end{split}\end{equation}
\begin{equation}\label{Eq:dmv}
\mathcal{L}_{\T}^\dmV = - \frac{1}{4}\frac{\alpha_\dmV}{\Lambda}~\r~\dmV_{\mu \nu} \dmV^{\mu \nu} + 2 \frac{\beta_\dmV}{\Lambda} ~\partial_\mu \i~\epsilon^{\mu \nu \rho \sigma} \dmV_\nu \partial_\rho \dmV_\sigma \, .
\end{equation}

We can make some remarks concerning the couplings of the moduli fields to the dark matter. First of all, similar to the case for the Higgs sector, the scalar dark matter does not couple to the imaginary part of the moduli. An immediate consequence, which will be discussed in what follows, is that only the real part of moduli contribute to the production of a scalar dark matter -- regardless of the Standard Model initial states.  

The interactions of fermions ($\Psi$, standard or dark) with moduli are essentially different from the scalar and vectorial cases, due to their chirality. The first aspect of this remark is evident from the amplitudes for their interactions with moduli:
\begin{align} \label{fermionamp}\begin{split}
{\cal M}_{\r \bar{\Psi} \Psi} &= - \frac{ i}{2\L} \bar{u}(p_1) (\slashed p_1 - \slashed p_2) (\a_\text{V}^\Psi - \a_\text{A}^\Psi \gamma_5) v(p_2) \\
&= - i \frac{\a_\text{V}^\Psi m_\Psi}{\L} \bar{u}(p_1) v(p_2)\, , \\
{\cal M}_{\i \bar{\Psi} \Psi} &= \frac{1}{2\L} \bar{u}(p_1) (\slashed p_1 + \slashed p_2) (\b_\text{V}^\Psi - \b_\text{A}^\Psi \gamma_5) v (p_2) \\
&= - \frac{\b_\text{A}^\Psi m_\Psi}{\L} \bar{u}(p_1) \gamma_5 v(p_2)\, ,
\end{split}\end{align}
where $p_1$ and $p_2$ are the four momenta of the fermions and $m_\Psi$ is the mass of the fermion. We notice that if the fermions are on-shell, we have an explicit dependence on their mass due to the Dirac equation. As a consequence, above the electroweak scale, standard fermions cannot produce any of the dark matter particles. The other aspect we point out is that the fermionic coupling to the real part of the moduli is CP-even, so that the corresponding rates will depend only on the vector coupling ($\a_\text{V}^\Psi$); and that the fermionic coupling to the imaginary part of the moduli is CP-odd and therefore the corresponding rates will depend only on the axionic coupling ($\b_\text{A}^\Psi$).

\section{Cosmological Moduli Problem}
\label{secIII}

Because of their gravitational interactions, moduli are long-lived fields. Thus one has to face two potential dangers. If the moduli lifetime is smaller than the age of our Universe, their decay might have released quite a large amount of entropy in the Universe which would dilute the contents of the Universe. On the other hand, if the moduli lifetime is larger than the age of our Universe, they might presently still be oscillating around the minimum of their potential. Thus, the energy stored in these oscillations may overclose the Universe. One refers to these problems as the cosmological moduli problem \cite{Coughlan:1983ci,*Banks:1993en,*deCarlos:1993wie,*Binetruy:2003dx}. Taken at their face values, such constraints dictates the moduli fields either to be super-light or to be super-heavy. We present here a short review on these well-known constraints, following Ref.~\cite{Coughlan:1983ci,*Banks:1993en,*deCarlos:1993wie,*Binetruy:2003dx}.

Using simple dimensional analysis the decay width of a modulus field $\T$ of mass $m_\T$ can be written as $\Gamma_\T \sim m_\T^3/M_P^2$. Since the age of the Universe is of order $H_0^{-1}$, where $H_0$ is the Hubble constant in the present Universe, we could infer that the modulus will decay at present times if $\Gamma_\T \sim H_0$, or in other words if its mass $m_\T$ is around $(H_0 M_P^2)^{1/3} \simeq 20$ MeV.

At first, we consider the case when $m_\T < 20$ MeV, that is when modulus has not decayed yet at the present time. As long as the Hubble is larger than the modulus mass, the friction term ($ 3 H\dot{\T}$) dominates in the equation of motion of the modulus field and the field $\T$ remains frozen at its initial value $f_\T$. When Hubble is of the order of the mass of the modulus or $H \sim m_\T$, that is when $T_I \sim \sqrt{m_\T M_P}$ (since we know that $H \sim T^2/M_P$), the field $\T$ starts oscillating around the minimum $\T_0$ of its potential. These coherent oscillations behave like non-relativistic matter thus the energy density of the modulus reads 
\begin{align}
	\rho_\T(T) = \rho_\T(T_I) \left(\frac{T}{T_I}\right)^3 \sim m_\T^2 f_\T^2 \left(\frac{T}{T_I}\right)^3 .
\end{align}
We know that the radiation energy density $\rho_\gamma(T)$ scales as $T^4$, thus $\rho_\T/\rho_\gamma$ scales as $1/T$. As the temperature of the Universe decreases the energy fraction stored in the moduli increases. Then there will be a time when the energy density of the modulus oscillations will dominate the total energy density of the Universe. Thus one needs to make sure that in the present Universe, $\rho_\T(T_0) < \rho_{c}$, where $T_0$ is the temperature of the present Universe and $\rho_{c}$ is the critical density. Using the above equation and $T_I = \sqrt{m_\T M_P}$, one can write this condition as
\begin{align}
	m_\T < M_P \left(\frac{\rho_c M_P}{f_\T^2 T_0^3}\right)^2 \sim 10^{-26}\  \mathrm{eV},
\end{align}
where we have assumed $f_\T \simeq M_P$. Thus, if $10^{-26}$ eV $< m_\T < 20$ MeV, there is huge amount of energy stored in the $\T$ field.

Next, we consider the case when $m_\T > 20$ MeV, that is when modulus field has already decayed at present times. Assuming the modulus field energy density dominates over the radiation energy density, the decay of the modulus occurs at a temperature $T_D$ when $H(T_D) \sim \Gamma_\T$, which can be re-expressed as 
\begin{align}
	\Gamma_\T^2 \sim \frac{\rho_\T(T_D)}{M_P^2}\, .
\end{align}
At the time of decay, all the energy density stored in the modulus is transferred into radiation energy density. Thus, the reheating temperature ($T_{\rh}$) is given by the condition $\rho_\T(T_D) \sim T^4_\rh$. Using the above equation we can obtain
\begin{align}
	T_\rh \sim \sqrt{M_P \Gamma_\T} \simeq \sqrt{\frac{m_\T^3}{M_P}}\,.
\end{align}
The entropy release due to the modulus decay must take place before the big-bang nucleosynthesis such that the abundance of the light elements is not affected. This condition, namely $T_\rh > 1$ MeV \cite{Kawasaki:1999na,*Kawasaki:2000en,*Hannestad:2004px}, gives $m_\T >10$ TeV. Thus for $20$ MeV $<m_\T <10$ TeV, the entropy release due to the decay of the modulus field would contradict with the current cosmological observations. In the absence of any other effects, cosmologically acceptable regime for the modulus mass is either super-light ($m_\T < 10^{-26}$ eV) or is super-heavy ($m_\T > 10$ TeV). This concludes the short review discussing the constraints on the moduli mass.

\section{Dark matter production}
\label{secIV}

The evolution of the dark matter number density $n_\dm$ is determined by the Boltzmann equation
\beq \label{Eq:dndt}
\frac{\text{d} n_\dm}{\text{d}t} = -3 H(t) n_\dm + R(T),
\eeq

where $H (t) = \frac{1}{\sqrt{3}M_P}  \sqrt{\rho_{tot}(t)}$ is the Hubble expansion rate, with $M_P \simeq 2.4 \times 10^{18}$ GeV the reduced Planck mass and $\rho_{tot}(t)$ the total energy density which changes with time, as we will see in what follows. $R(T)= n_{SM}^2\langle \sigma v \rangle_{\sm \sm \rightarrow \dm \dm}$ is the production rate (number of dark matter particles produced per unit of time and unit of volume, see Appendix~\ref{Ap:SqAmpRate} for details) which, for a process $(1,2 \rightarrow 3,4)$, is given  by
\beq R(T) = \int f_1 f_2 \frac{E_1 E_2 \text{d}E_1 \text{d}E_2
  ~\text{d}\cos \theta_{12}}{1024 \pi^6} \int |{\cal M}|_i^2 \text{d}
\Omega_{13} \,,
\label{Eq:rt}
\eeq
with $E_i$ and $f_i$ being the energy and the (thermal) distribution function of the particle $i$ respectively, $\theta_{12}$ being the angle between the incoming particles 1 and 2 in the laboratory frame, and $\Omega_{13}$ is the solid angle between the outgoing particles 1 and 3 in the rest frame.

Since we are interested in mediators with masses that could be near the reheating scale, the contribution to the total energy density of the inflaton field (labeled by $\phi$) may dominate over the contribution of radiation (labeled by $\gamma$) in the Hubble rate. In order to find the correct amount of dark matter, Eq.\eqref{Eq:dndt} needs to be solved numerically along with the following coupled equations (see for instance \cite{giudice_largest_2000,garcia_enhancement_2017}):
\begin{align}\label{Eq:setboltzmann}
\frac{\text{d}\rho_\gamma}{\text{d}t} &= -4H\,\rho_\gamma+\Gamma_\phi\,\rho_\phi+2\langle\sigma v\rangle\langle E_\dm\rangle\left[n_\dm^2-(n_\dm^\text{eq})^2\right]\, , \notag \\ 
\frac{\text{d}\rho_\phi}{\text{d}t} &= -3H\,\rho_\phi-\Gamma_\phi\,\rho_\phi\, ,
\end{align}
with $\langle E_\dm \rangle$ the mean energy of the dark matter and  $\Gamma_\phi$ the inflaton total decay width. We have defined $T_\rh$ as the temperature in a radiation dominated Universe after the inflaton decay, $\Gamma_\phi = H(T_\rh)$.

In the present work, we have solved the set of three coupled differential equations above, but it is instructive to find analytic solutions for the limiting cases of inflaton and radiation domination. In fact, we have checked that this is a good approximation, since there is a recognizable change of regime in the Hubble rate near the reheating temperature (see for instance \cite{mazumdar_quantifying_2014}). In the radiation dominated era, we can use the familiar relations\footnote{For simplicity, we will not assume changes in the energetic and entropic relativistic degrees of freedom, $g_e$ and $g_s$ respectively. In what follows, we have set $g_e = g_s = 106.75$.}
\begin{equation}
\frac{\text{d}}{\text{d}t} = - H(T) T
\frac{\text{d}}{\text{d}T} \quad \textrm{with}~~ H(T) =
\sqrt{\frac{g_e}{90}} \pi \frac{T^2}{M_P},
\end{equation}
while in the inflaton dominated era, we have \cite{bhattacharyya_freezing-dark_2018}
\begin{equation}
\frac{\text{d}}{\text{d}t} = - \frac{3}{8} H(T) T
\frac{\text{d}}{\text{d}T} \quad \textrm{with}~~ H(T) =
\sqrt{\frac{5 g_\max^2}{72 g_\rh}} \pi \frac{T^4}{T_\rh^2 M_P},
\end{equation}
where $g_\rh$ and $g_\max$ are respectively the relativistic degrees of freedom at reheating temperature after inflation and at the maximal temperature reached during the reheating process, $T_\max$.

The dark matter relic density, $\Omega h^2 \equiv m_\dm n_\dm/\rho_{c}$ with $\rho_{c}$ the critical density today, may be split into a radiation dominated (RD) and an inflaton dominated (ID) contributions:
\begin{equation}\begin{split}\label{relicsplit}
\O h^2 &\cong \O h^2_{RD} + \O h^2_{ID} \\
& \sim 4 \times 10^{24}~m_\dm \left( \int_{T_0}^{T_\rh} dT \frac{R(T)}{T^6} \right. \\
&~~~~~~~~~~~~~~~~~~~~~~~ \left. +\, 1.07 ~T_\rh^7 \int_{T_\rh}^{T_\max} dT \frac{R(T)}{T^{13}} \right) \\
& \equiv \O h^2_{RD} B_F,   
\end{split}\end{equation}
where $T_0$ is the temperature of the present Universe. In the above equation we have defined a ``boost factor''  $B_F$ ($= 1 + \frac{\O h^2_{ID}}{\O h^2_{RD}}$), which quantifies the fraction of dark matter produced during the ID era to the amount of dark matter produced during the RD era. Notice that the fraction of dark matter produced during the reheating stage is $(B_F-1)/B_F$. 

From this expression we can see that the production of dark matter during reheating ($T_\rh < T < T_\max$) might be relevant if the temperature dependence in the rate is sufficiently high. Let us parametrize the rate as $R(T) \propto T^n$ and denote the boost factor as $B_F^{(n)}$. For $n \geq 12$, dark matter production during reheating is comparable or dominant over production during radiation dominated era. For instance, for $T_\max = 100\, T_\rh$ we find 
\begin{equation}
B_F^{(10)} \approx 1 + 1.07 \times \frac{5}{2}  \left(1 + \frac{T_\rh^2}{T_\max^2} \right) \simeq 3.68 \, , 
\end{equation}
\begin{equation}
B_F^{(12)} \approx 1 + 1.07 \times 7 \ln\left(\frac{T_\max}{T_\rh}\right) \simeq 35.5 \, , 
\end{equation}
and, for $n>12$,
\begin{equation}
B_F^{(n>12)} \approx 1 + 1.07 \times \frac{n-5}{n-12} \left(\frac{T_\max}{T_\rh}\right)^{n-12}\, .
\end{equation}
 The percentage of dark matter production during reheating for $n=10$ is therefore of $\sim 73\%$ whereas for $n=12$ if of $\sim 97\%$. An important point we want to emphasize here is that if the mediators between the dark and visible sectors are close to the reheating scale and if the production rate of dark matter have a high temperature dependence, we cannot avoid the contribution of the inflaton to the Hubble rate.

\section{Results and discussion}
\label{secV}
\subsection{Production rate}

The squared amplitudes responsible for the model-dependent behavior of the production rates of three different dark matter candidates are given in Appendix \ref{Ap:SqAmpRate}\footnote{In our analysis, we concentrate on the first possibility discussed for the mass parameter of the Higgs (see Eq.\eqref{caseA}). For the possibility where the mass parameter is given by Eq.\eqref{caseB} the result is similar to the case in Eq.\eqref{caseA}.}. Before presenting the exact solution of the rate (see Eq.\eqref{eq:exactrate}), we recognize from the squared amplitudes three regimes which depend on the relation between the mass of the mediators and the temperature of the thermal bath:

\begin{itemize}
\item \textit{the light regime} -- when the mediator mass is much below the temperature of the thermal bath $T$ ($m_{\r, \i} \ll T $);
\item \textit{the pole regime} -- when the mediator mass is of the same order than the temperature ($m_{\r, \i} \sim T$), where we might use the narrow width approximation;
\item \textit{the heavy regime} -- when the mediator mass is much above the temperature ($m_{\r, \i} \gg T$) .
\end{itemize}

Far from the pole of the propagator, we might assume $\G_{\r, \i} \ll m_{\r, \i}$. In the limit $m_\dm \ll T$, we can obtain analytic solutions for the production rate $R_{s_f}^j$ of a dark matter of spin $s_f$ due to the exchange of a mediator $j$:

\begin{equation}\label{rate01}
R_{0,1}^j (T) = \delta^j_{0,1} \times \left\{ \begin{array}{lc}
\dfrac{T^8}{\L^4}\, , ~~~~~~~~~~~~~~~~~~~~~~ (m_j \ll T) \\ \noalign{\medskip} \dfrac{m_j^8}{\L^4} \dfrac{T}{\G_j} K_1\Big(\dfrac{m_j}{T}\Big)\, , ~~~~~~ (m_j \sim T) \\ \noalign{\medskip} 
\dfrac{T^{12}}{m_j^4\L^4}\, , ~~~~~~~~~~~~~~~~~~ (m_j \gg T) 
\end{array} \right.
\end{equation}

\begin{equation}\label{rate1o2}
R_{1/2}^j (T) = \delta^j_{1/2} \times \left\{ 
\begin{array}{lc}
\dfrac{m_{\dm}^2 T^6}{\L^4}\, , ~~~~~~~~~~~~~~~~~~~~ (m_j \ll T) \\ \noalign{\medskip} 
\dfrac{m_\dm^2 m_j^6}{\L^4} \dfrac{T}{\G_j} K_1\Big(\dfrac{m_j}{T}\Big)\, , ~~~~ (m_j \sim T) \\ \noalign{\medskip} 
\dfrac{m_\dm^2 T^{10}}{m_j^4\L^4}\, , ~~~~~~~~~~~~~~~~~~~ (m_j \gg T) 
\end{array} \right. 
\end{equation}
where the proportionality constants $\delta^j_{s_f}$ are given in Table ~\ref{tab:rates}.

We have computed numerically the total production rates, Eq.\eqref{eq:exactrate}, where the integration was computed using the CUBA package \cite{hahn_cuba_2005}, with Bose-Einstein distribution function for the Higgs and gauge bosons in the initial states. 

\begin{figure}[t!]
\centering
\includegraphics[width=0.5\textwidth]{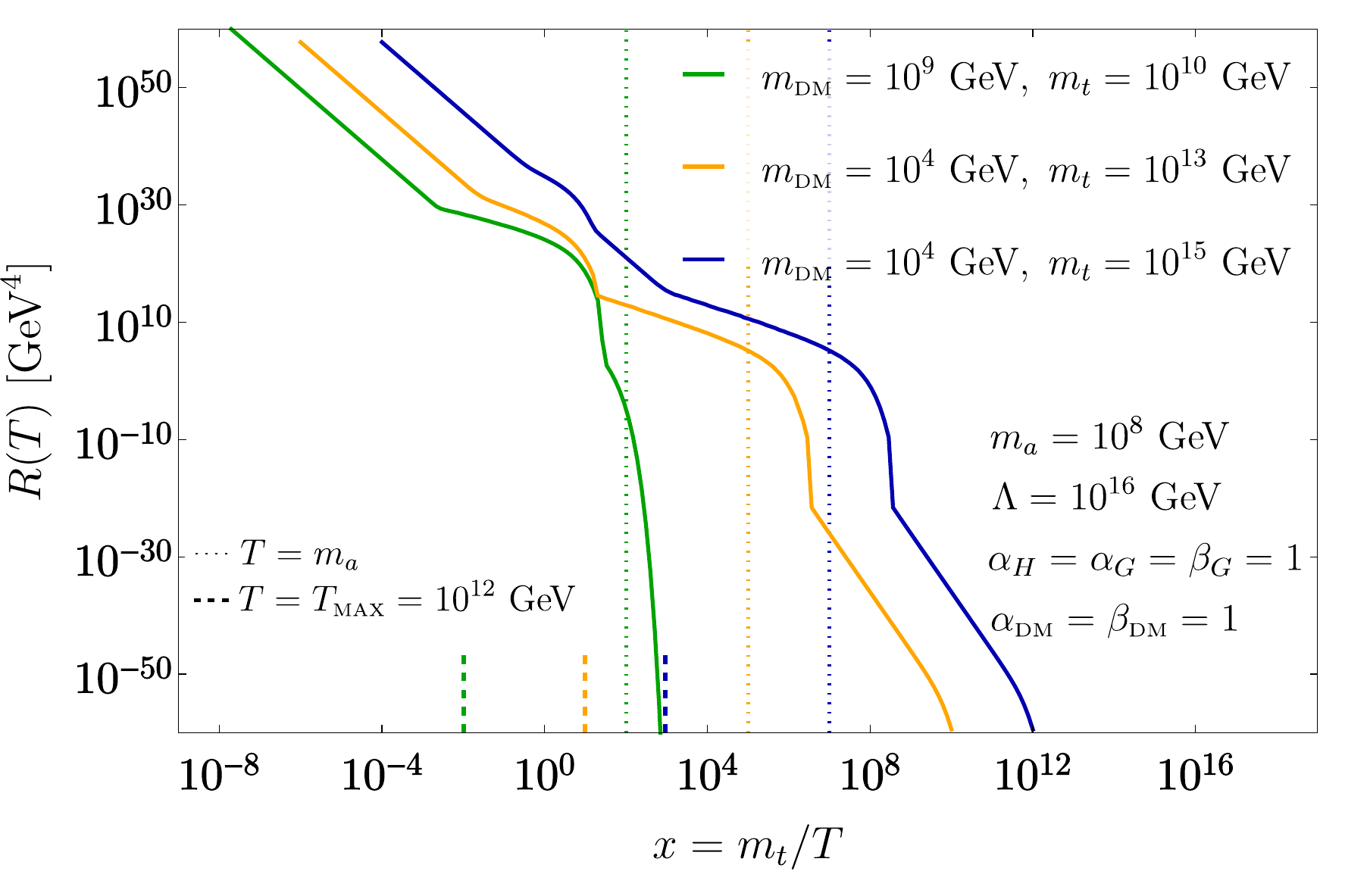}
\caption{\small Evolution of the production rate of fermionic dark matter as function of the temperature for different masses of dark matter and real component of the modulus field.}
\label{Fig:rate}
\end{figure}

In Fig.~\ref{Fig:rate}, we show the exact solutions of the total production rate of the fermionic dark matter for a representative set of free parameters, as a function of the variable $x = m_s/T$ which may be regarded as a parametrization of time. We set the new physics scale $\Lambda$ to be $10^{16}$~GeV  (GUT scale), $T_{\max}=10^{12}$ GeV and the mass of the axionic modulus to be $10^8$~GeV. For simplicity, all the couplings are set to unity. From left to right, the mass of the real component of the modulus is set to $10^{10}$, $10^{13}$ and $10^{15}$~GeV (green, orange and blue curves, respectively). The mass of the fermionic dark matter is set to be between the mediator masses in the first case ($10^9$~GeV) and to be relatively light in the second and third cases ($10^4$~GeV).

It is easy to understand the mechanism at work in the dark matter production after a look at Fig.~\ref{Fig:rate}. First of all, a general feature of the rate is the strong temperature dependence: the higher the temperature (small $x$ region), the more dark matter would be produced. The second generic feature is the threshold for dark matter production which is due to the Boltzmann suppressed photon distribution having $T>M_{DM}$ (large $x$). This happens just after $x=10$, $10^9$ and $10^{11}$ for the three case respectively.

Between those two extremes, we can notice the effects of the pole regions once $T$ reaches $\mre$ ($x \sim 1$) and $\mim$ ($x=10^2$, $10^5$ and $10^7$ for $\mre=10^{10}$, $10^{13}$ and $10^{15}$ GeV respectively). Notice that the production rates for the scalar dark matter would not have the effect of the poles of $\mim$ since it couples only with the real component of the modulus. The production rate of a vectorial dark matter would have the same qualitative features of the fermionic case but with a steeper bend at high temperatures, since the temperature dependence in the heavy regime is $T^{12}$ in the vector dark matter case and $T^{10}$ in the fermionic case.

The presence of the pole regions depend on the low and high temperature thresholds. It will not appear if the Boltzmann suppression takes place before it (as in the green curve, for $x \sim 100$). Since the Universe has a maximal temperature, fixed to $10^{12}$ GeV in Fig.~\ref{Fig:rate}, the production rate will have maximal values at $x = 10^{-2}, 10$ and $10^3$. As a consequence, the pole due to the real component exchange would not contribute for the cases in orange and blue.

\subsection{Relic abundance}

From the approximate rates given in the last section, we can have an idea about the parameter space in agreement with the inferred value of the dark matter relic density $\Omega h^2 = 0.1200 \pm 0.0012 $ \cite{planck_collaboration_planck_2018}. Taking the limit of heavy moduli, we find
\begin{widetext}
\begin{equation}
\frac{\Omega h^2}{0.12} \approx 
\begin{cases}
\frac{B^{(12)}_F}{35.5} \a_S^2\frac{\a_\sm^2}{5} \Big(\frac{m_\dm}{1.2\times 10^{14}\gev}\Big)\Big(\frac{T_\rh}{10^{10}\gev}\Big)^7 \Big(\frac{\Lambda}{10^{15}\gev}\Big)^{-4} \Big(\frac{\mre}{10^{13}\gev}\Big)^{-4}\, ,& \mathrm{(scalar ~DM)} \\
\frac{B^{(10)}_F}{3.68} \Big(\frac{m_\dm}{3.2\times 10^{10}\gev}\Big)^3 \Big(\frac{T_\rh}{10^{10}\gev}\Big)^5 \Big(\frac{\Lambda}{10^{15}\gev}\Big)^{-4} \left[\frac{\left(\a_\text{V}^\dmF\right)^2}{2}\frac{\a_\sm^2}{5}\Big(\frac{\mre}{10^{13}\gev}\Big)^{-4} + \frac{\left(\b_\text{A}^\dmF\right)^2}{2}\b_G^2 \Big(\frac{\mim}{10^{12}\gev}\Big)^{-4} \right]\, ,& \mathrm{(fermionic ~ DM)} \\
\frac{B^{(12)}_F}{35.5} \Big(\frac{m_\dm}{1.5\times 10^{12}\gev}\Big) \Big(\frac{T_\rh}{10^{10}\gev}\Big)^7 \Big(\frac{\Lambda}{10^{15}\gev}\Big)^{-4} \left[\frac{\a_{\dmV}^2}{2}\frac{\a_\sm^2}{25}\Big(\frac{\mre}{10^{13}\gev}\Big)^{-4} + \frac{\b_{\dmV}^2}{2}\b_G^2 \Big(\frac{\mim}{10^{12}\gev}\Big)^{-4} \right]\, .& \mathrm{(vectorial ~ DM)}
\end{cases}
\label{Eq:omega}
\end{equation}
\end{widetext}
It is important to underline that the expressions in Eq.\eqref{Eq:omega} are computed with simplified hypothesis, especially in the limit $\mre \gg T_\max$. Comparing Eq.\eqref{Eq:omega} with our numerical results, we noticed that pole effects due to the exchange of $\r$ can be important even when $m_\dm$ lies above $T_\max$ as the enhancement due to a small value width can compensate the Boltzmann suppression $e^{-m_\dm / T_\max}$.

\begin{figure}[h!]
\centering
\includegraphics[scale=0.55]{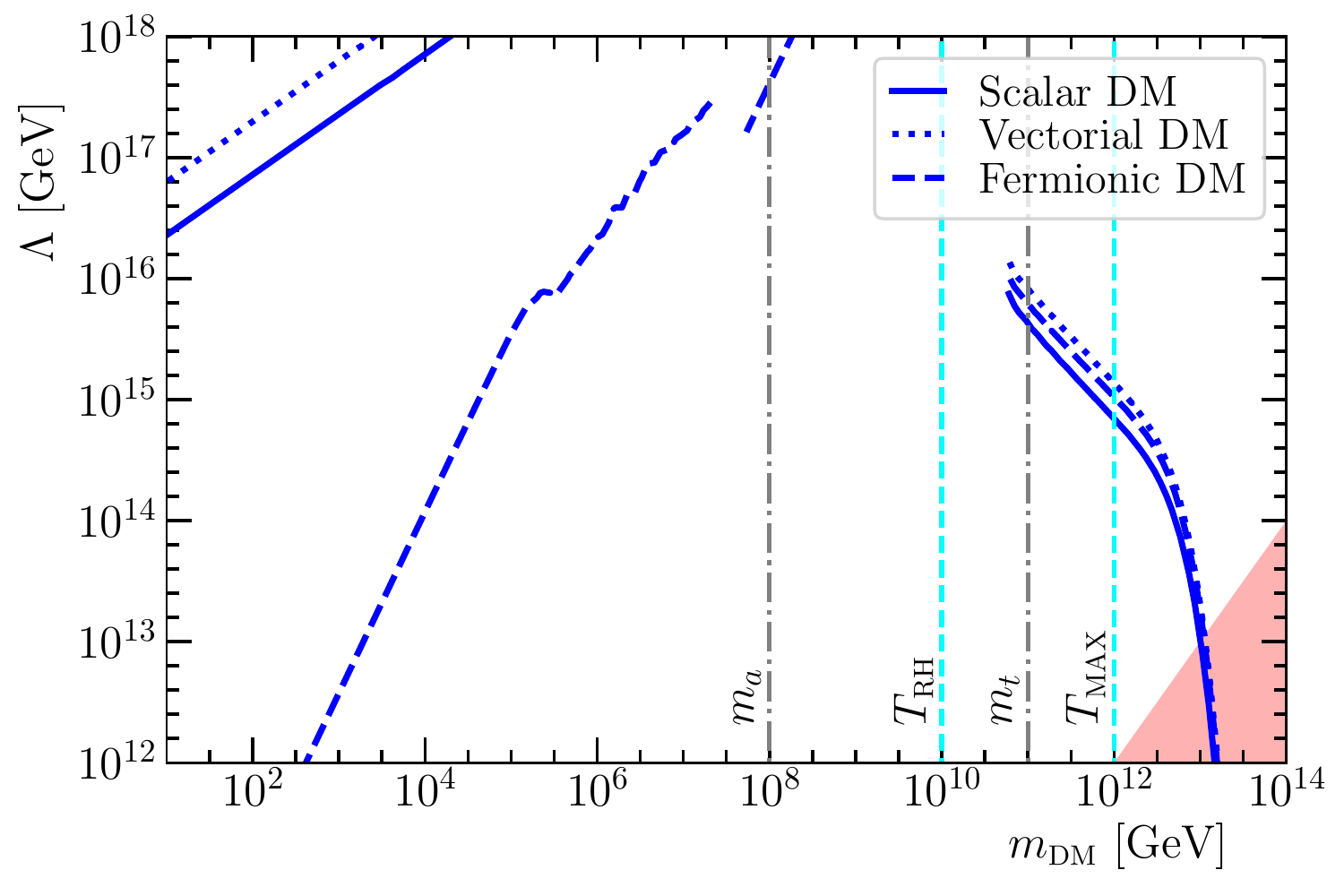}
\caption{\small Contours respecting $\Omega h^2 = 0.12$ in the ($m_\dm$, $\L$) plane for real and imaginary parts of modulus with masses $\mre = 10^{11}$ GeV and $\mim = 10^{8}$ GeV, respectively. For an illustrative purpose, we set $T_\rh = 10^{10}$ GeV, $T_\max = 100\, T_\rh$ and all couplings are set to unity. The region in red is not reliable since $\L < m_\dm$.}
\label{Fig:mdm-lam}
\end{figure}

\begin{figure}[h!]
\centering
\includegraphics[scale=0.55]{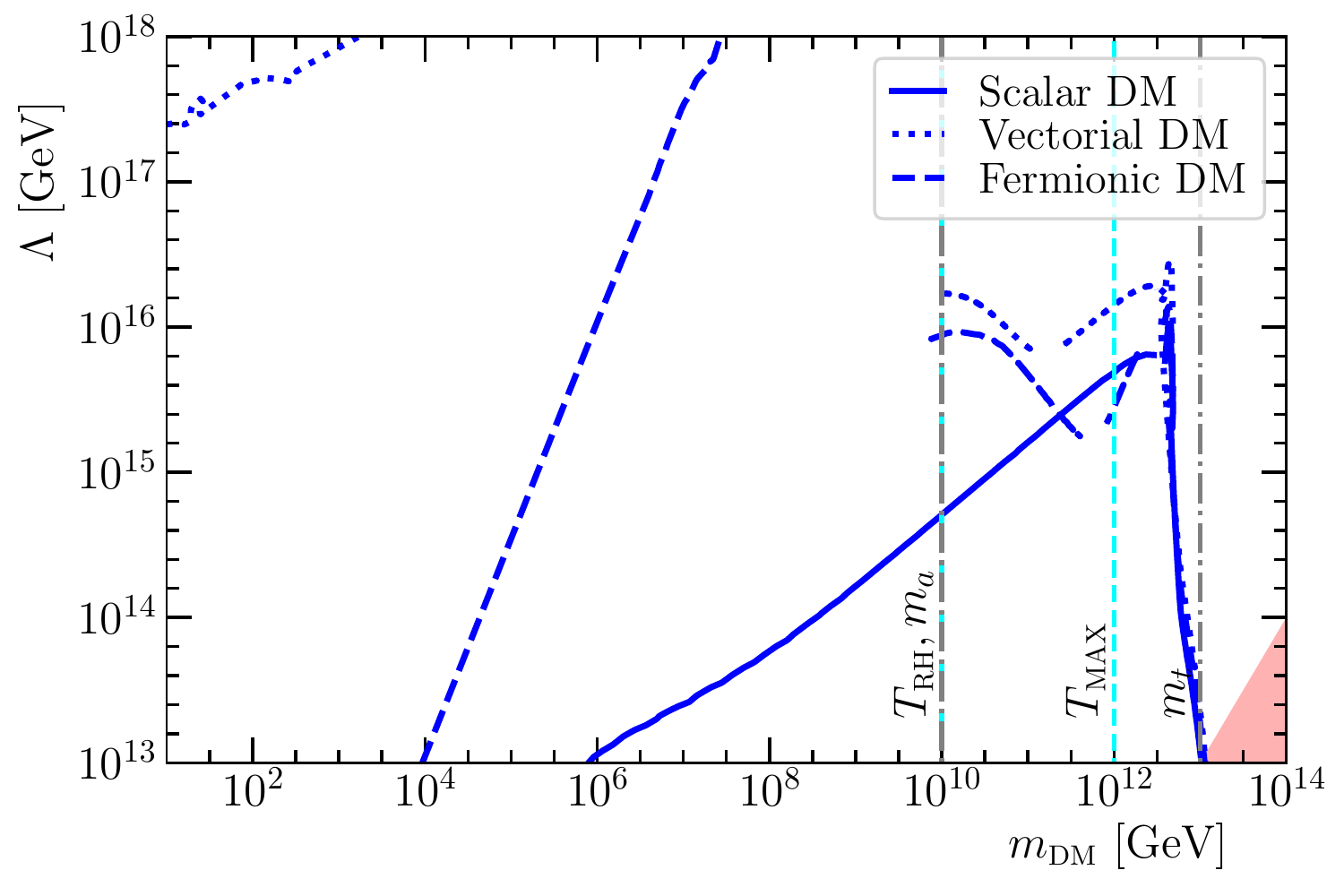}
\caption{\small Contours respecting $\Omega h^2 = 0.12$ in the ($m_\dm$, $\L$) plane for all parameters set as in Fig.~\ref{Fig:mdm-lam} but for heavier moduli: $\mim = 10^{10}$ GeV and $\mre = 10^{13}$ GeV.}
\label{Fig:mdm-lam-13}
\end{figure}

Our results are summarized in Figs.~\ref{Fig:mdm-lam} and \ref{Fig:mdm-lam-13}, where we have plotted the contours in the parameter space $(m_\dm,\L)$ corresponding to $\Omega h^2 = 0.12$ for the scalar, fermionic and vectorial dark matter (solid, dotted and dashed blue lines, respectively). For simplicity, all the couplings in the Lagrangian are set to unity. The split between the reheating and maximal temperatures depend, of course, on the inflationary model. We set this split to be $T_\max = 100\, T_\rh$ inspired by a Starobinsky-like potential \cite{starobinsky_new_1980}, which sets the initial condition for the inflaton energy density (after the end of inflation) to be $\rho_{end} \simeq 0.175\, m_\phi^2 M_P^2$ \cite{ellis_post-inflationary_2015}, for an inflaton mass of $m_\phi = 10^{13}$ GeV and for an inflaton decay width $\Gamma_\phi = \a_\phi^2 m_\phi$ with $\a_\phi = 10^{-8}$. We set the reheating scale to be $10^{10}$ GeV. In Fig.~\ref{Fig:mdm-lam} we have set $\mim = 10^{8}$ GeV and $\mre = 10^{11}$ GeV, whereas we explore a scenario with heavier mediators in Fig.~\ref{Fig:mdm-lam-13}, with $\mim = 10^{10}$ GeV and $\mre = 10^{13}$ GeV. With this set of parameters, it is imperative to consider the presence of the inflaton energy density in the Hubble rate. Indeed, our complete analysis was performed by solving the complete set of Boltzmann equations, and integrating over the whole phase space with Bose-Einstein distribution functions for the initial states (since standard fermions do not contribute for the dark matter production in our scenario).

The first thing the reader might notice in each figure is the stronger dependence of the fermionic contour on the dark matter mass, as compared to the scalar and vectorial cases which have a linear dark matter mass dependence coming from the relic density definition. This is easy to understand since the amplitude for the annihilation of the standard model states into fermionic dark matter depend explicitly on the dark matter mass (see Eq.\eqref{fermionamp}). It is therefore easier to see in the fermionic contours the following generic feature: the lighter the dark matter, the smaller the new physics scale for the same relic density value. On the other hand, the contours of the scalar and vectorial dark matter have similar behavior, as suggested by the approximate expressions for the rate in Eq.\eqref{rate01}. Comparing the scalar and vectorial cases in each figure, we see that for any dark matter mass, the same relic density value is achieved for larger values of $\L$ in the vectorial case, since a vectorial dark matter receives contribution from the imaginary part of modulus and the rate is therefore higher -- which is not the case for the scalar dark matter. This feature is more accentuated in Fig.~\ref{Fig:mdm-lam-13}, as we will be able to understand in what follows.

In the parameter region where $m_{\r,\i} > 2 m_\dm$, the mediators can decay on-shell to the dark matter particles whenever the pole can be reached ($m_{\r,\i} < T_\max$). The dark matter production in this region of the parameter space is therefore enhanced and we can understand that by increasing $m_\dm$ the $\Lambda$ needs to be increased as well to provide the same relic density. So, the contour of $\Omega h^2 = 0.12$ monotonically increases as $m_\dm$ is increased until the threshold for on-shell production is reached, for $m_\dm \simeq m_{\r,\i}/2$. In the parameter region with $m_\dm > m_{\r,\i}/2$, the dark matter is produced from the Standard Model through off-shell mediators. Thus in this regime, the cross-section and the rate are much lower compared to the pole-enhanced region. As the rate is much lower in this regime, the effective scale ($\Lambda$) needs to be also decreased to much lower values so as to satisfy the Planck constraint on the dark matter relic density. This is why we can observe a very sharp transition between the on-shell and off-shell production regimes. This decreasing in $\Lambda$ continues as we increase the dark matter mass, and becomes drastically accentuated because of the Boltzmann suppression in the rate, as we saw in Fig.~\ref{Fig:rate}. The heavier the dark matter, the less its production, and then a smaller $\Lambda$ is needed to compensate the loss of production. The process goes on until the point where it becomes impossible to produce dark matter from the Standard Model. In fact, if dark matter is that heavy, our effective theory approach is no longer in a firm footing since we would enter into a unreliable region of our parameter space, with $\L < m_\dm$ (red shaded region).

Because of their accentuated dark matter mass dependence, the fermionic dark matter contours allow much lower effective scale values for the entire region where dark matter is produced through the on-shell decay of the mediators. As the dark matter mass increases to $m_\dm \gtrsim m_{\r,\i}/2$ the off-shell production becomes the dominant one and the rate slowly merges to the rate of the other two dark matter cases. As a result of it, in this regime the allowed contour for fermionic dark matter mimics the scalar and vectorial dark matter curve. Notice that in the fermionic contour we see a kink around $m_\dm \simeq m_a/2$\footnote{For the vectorial case, this kink would happen for unrealistic high values of $\L$, which are not displayed in our results.}, this is due to the fact that, before this region, the fermionic dark matter production was due to on-shell decay of both the real and imaginary part of the moduli (pole region). As the dark matter mass reaches $m_\dm \simeq m_a/2$, the dark matter production through imaginary part of the moduli becomes off-shell while production through the real part of the moduli remains on-shell. At this point, the dark matter production rate reduces and as a result there is a dip in the curve to compensate this change. Above this regime, the slope of the curve changes as the dominant contribution to the rate is only through the on-shell exchange of the real part of the moduli.

Finally, we can understand the main difference between Figs.~\ref{Fig:mdm-lam} and \ref{Fig:mdm-lam-13}. Heavier mediators lead to suppressed rates, which brings the effective scale to lower, and in our case more reasonable, values. In the fermionic and vectorial dark matter contours, we observe in Fig.~\ref{Fig:mdm-lam-13} viable regions which could not be present in Fig.~\ref{Fig:mdm-lam}, corresponding to the combination of off-shell production from imaginary modulus and on-shell production of real modulus that happens for dark matter masses between $T_\rh$ and $T_\max$. Concerning the scalar contour, we see in Fig.~\ref{Fig:mdm-lam-13} that the enhancement of the on-shell production from the real modulus (the only mediator possible in this case) is not more efficient than the suppression due to the exchange of a very heavy modulus.  

We finalize our discussion with the following lesson for the reader. The derivative couplings of the operators connecting the visible and dark sectors, due to moduli field exchanges, generate high temperature dependence in the production rates. As a consequence, the dark matter candidates considered in the present work are mainly produced during the reheating process in the Early Universe. Additionally, the pole enhancements inter-played by the moduli mediators control the regions of the parameter space which can account for the right amount of dark matter in the Universe.

\section{Conclusion}
\label{secVI}

We showed in this work that moduli fields which are present in several extensions of the Standard Model, even if very heavy, can play the role of a mediator between the dark sector and the Standard Model. Through its couplings to the thermal bath, dark matter can be produced at a sufficient rate to fulfill the observed relic abundance. 
The main production takes place through an ultra-violet freeze-in mechanism where the majority of dark matter is created at the early stage of reheating. Our main results are summarized in Figs.~\ref{Fig:mdm-lam} and \ref{Fig:mdm-lam-13}, where the correct relic abundance can be obtained for a broad range of dark matter mass and an effective scale $\Lambda$ around the unification/string scale. Finally, our results are quite general, and can be applied to any ultra-violet models where moduli fields are present.

\begin{acknowledgments}
The authors want to thank especially W.~Buchmüller and K.~Olive for detailed and insightful discussions. This research has been supported by the (Indo-French) CEFIPRA/IFCPAR Project No.~5404-2. Support from CNRS LIA-THEP and the INFRE-HEPNET of CEFIPRA/IFCPAR is also acknowledged. D.C. would like to thank CERN theory group for their hospitality and support. E.D. was supported in part by the ANR grant Black-dS-String ANR-16-CE31-0004-01.  M.D. acknowledges support from the Brazilian PhD program ``Ci\^encias sem Fronteiras''-CNPQ Process No. 202055/2015-9. This work was also supported by the France-US PICS no.~06482, PICS MicroDark. This project has received funding/support from the European Union’s Horizon 2020 research and innovation programme under the Marie Sklodowska-Curie: RISE InvisiblesPlus (grant agreement No.~690575) and the ITN Elusives (grant agreement No.~674896).

\end{acknowledgments}

\appendix

\section{Width of the moduli fields}

Here we provide the expressions for the decay widths of the moduli fields\footnote{Effects of finite temperature on the decay width of moduli, as discussed for instance in \cite{bodeker_moduli_2006}, will not be relevant for us since we will not consider the moduli as ultra-relativistic species in our analysis.}. The real components of the moduli may decay into scalars and vectors of the Standard Model, since the decay into fermions are not allowed above the electroweak symmetry breaking. We have therefore
\begin{equation}\begin{split}
    \Gamma_\r &= 4~\G_{\r \to HH} + 12~\G_{\r \to GG} + \G_{\r \to \dm \dm}\\
&= \frac{\mre^3}{\pi \L^2} \Big( \frac{f_H(\mre^2)^2}{8} \sqrt{1-\frac{4 \m_0^2}{\mre^2}} + \frac{3}{16} \a_\ttiny{G}^2 \Big) + \Gamma_{\r \to \dm \dm},
\end{split}\end{equation}

The distinct ways of writing the mass parameter of the Higgs lead us to define the function
\begin{equation}
f_H (x) \equiv \left\{
\begin{array}{lc}
\a_H ~~~~~~~~~~~~~~~~~~~~~~~~~~~~~~~ \text{(case of Eq.\ref{caseA})} \\ \noalign{\medskip}
 \a_H \Big( 1-\frac{2\m_0^2}{x}\Big) + \frac{2\L}{x} \frac{\langle F \rangle}{M_P} ~~~~~~ \text{(case of Eq.\ref{caseB})} 
\end{array}
\right. 
\end{equation}

In the case of the imaginary component, decay into scalars is prohibited, and we have simply
\begin{equation}\begin{split}
    \Gamma_\i &= 12~\G_{\i \to GG} + \G_{\i \to \dm \dm}\\
&= \frac{\mim^3}{\pi \L^2} 3 \b_\ttiny{G}^2 + \Gamma_{\i \to \dm \dm}.
\end{split}\end{equation}

The partial decay widths of the real and imaginary parts of the modulus into dark matter read respectively
\begin{equation}\begin{split}
&\Gamma_{\r \to \dm \dm} \\
& = \frac{\mre^3}{\pi \L^2} \sqrt{1-\frac{4 m_\dm^2}{\mre^2}} \times 
\left\{ \begin{array}{lc}
\frac{\a_\dmS^2}{32} \Big( 1 - \frac{2m_\dm^2}{\mre^2}\Big)^2,~~~~~~~~~~~\text{for $\dmS$} \\ \noalign{\medskip} 
\frac{(\a_V^\dmF)^2}{8} \frac{m_\dm^2}{\mre^2} \Big( 1 - \frac{4m_\dm^2}{\mre^2}\Big),~~\text{for $\dmF$} \\ \noalign{\medskip} \frac{\b_\dmV^2}{64}\Big( 1 - \frac{4m_\dm^2}{\mre^2} + \frac{6m_\dm^4}{\mre^4} \Big),~\text{for $\dmV$} \\ \noalign{\medskip} \end{array} \right.
\end{split}\end{equation}
and
\begin{equation}\begin{split}
& \G_{\i \to \dm \dm}\\
& = \frac{\mim^3}{\pi\L^2} \sqrt{1-\frac{4 m_\dm^2}{\mim^2}} \times 
\left\{ \begin{array}{lc}
0,~~~~~~~~~~~~~~~~~~~~~~~~~~~~~~~\text{for $\dmS$} \\\frac{1}{8}(\b_\dmF \b_5^\dmF)^2 \frac{m_\dm^2}{\mim^2},~~~~~~~~~~~~~\text{for $\dmF$} \\ \noalign{\medskip} 
\frac{\b_\dmV^2}{4}\Big( 1 - \frac{4m_\dm^2}{\mim^2} \Big),~~~~~~~~~~~~\text{for $\dmV$} \\ \noalign{\medskip} \end{array} \right.
\end{split}\end{equation}

\section{Squared amplitudes}\label{Ap:SqAmpRate}

\begin{table*}[t]
\centering
 \begin{tabular}{|l||*{3}{c|}}\hline
\backslashbox{$SM$}{$DM$}
&\makebox[3em]{spin-0}&\makebox[4em]{spin-1/2}&\makebox[3em]{spin-1}\\\hline\hline
spin-0 & $ f_H^2 (s)$ & $2 f_H^2 (s)$ & $1/2~ f_H^2 (s)$   \\\hline
spin-1/2 & $0$ & $0$ & $0$ \\\hline
spin-1 & $1/2~\a_G^2$ & $ \a_G^2$ & $ 1/4~ \a_G^2$ \\\hline
\end{tabular}
\hspace{1cm}
\begin{tabular}{|l||*{3}{c|}}\hline
\backslashbox{$SM$}{$DM$}
&\makebox[3em]{spin-0}&\makebox[4em]{spin-1/2}&\makebox[3em]{spin-1}\\\hline\hline
spin-0 & $0$ & $0$ & $0$   \\\hline
spin-1/2 & $0$ & $0$ & $0$ \\\hline
spin-1 & $0$ & $ 16 \b_G^2$ & $ 64 \b_G^2$ \\\hline
\end{tabular}
\caption{Coefficients of the squared amplitudes: $\l_{s_i,s_f}^t$ (left) and $\l_{s_i,s_f}^a$ (right) (Eqs. \ref{M0}, \ref{M12} and \ref{M1}).}\label{tab:amp}
\end{table*}

\begin{table*}[t]
\centering
\begin{tabular}{|l||*{3}{c|}}\hline
\backslashbox{regimes\\of $\r$}{$DM$}
&\makebox[3em]{spin-0}&\makebox[2em]{spin-1/2}&\makebox[2em]{spin-1}\\\hline\hline
Light & $\frac{\pi^3}{108000} \a_\dmS^2 \a_\sm^2$ & $\frac{\zeta(3)^2}{8\pi^5} {(\a_V^\dmF)}^2 \a_\sm^2$ & $\frac{\pi^3}{21600} \a_\dmV^2 \a_\sm^2$   \\\hline
NWA & $\frac{1}{1024\pi^4} \a_\dmS^2 \a_\sm^2$ & $\frac{1}{256\pi^4} {(\a_V^\dmF)}^2 \a_\sm^2$ & $\frac{1}{2048\pi^4} \a_\dmV^2 \a_\sm^2$ \\\hline
Heavy & $\frac{64\pi^7}{19845} \a_\dmS^2 \a_\sm^2$ & $ \frac{72\zeta(5)^2}{\pi^5} {(\a_V^\dmF)}^2 \a_\sm^2$ & $ \frac{32\pi^7}{19845} \a_\dmV^2 \a_\sm^2$ \\\hline
\end{tabular}\hspace{.1cm}
\begin{tabular}{|l||*{3}{c|}}\hline
\backslashbox{regimes\\of $\i$}{$DM$}
&\makebox[3em]{spin-0}&\makebox[3em]{spin-1/2}&\makebox[3em]{spin-1}\\\hline\hline
Light & $0$ & $\frac{6\zeta(3)^2}{\pi^5} {(\b_A^\dmF)}^2 \b_G^2$ & $\frac{8\pi^3}{225} \b_\dmV^2 \b_G^2$   \\\hline
NWA & $0$ & $\frac{3}{16\pi^4} {(\b_A^\dmF)}^2 \b_G^2$ & $\frac{3}{8\pi^4} \b_\dmV^2 \b_G^2$ \\\hline
Heavy & $0$ & $\frac{3456\zeta(5)^2}{\pi^5} {(\b_A^\dmF)}^2 \b_G^2$ & $ \frac{8192\pi^7}{6615} \b_\dmV^2 \b_G^2$ \\\hline
\end{tabular}
\caption{Coefficients appearing the approximate rates (Eqs. \ref{rate01} and \ref{rate1o2}): $\delta_{s_f}^\r$ (left) and $\delta_{s_f}^\i$ (right). We have defined $\a_\sm^2 \equiv 2\a_H^2+3\a_G^2$, since we assume $\m_0^2 \ll s$. Except for the NWA cases, we have used Bose-Einstein statistics for the initial state distribution functions.}\label{tab:rates}
\end{table*}

In what follows, we provide the expressions for the squared amplitudes of $s$-channel SM annihilations into DM candidates of spin $s_f$ ($|\M|_{s_f}^2$):
\begin{equation}\label{M0}
|\M|^2_0 = \frac{\a_\dmS^2}{\L^4} \frac{ s^4 \Big( 1 - \frac{2m_\dm^2}{s} \Big)^2}{(s-\mre^2)^2+\mre^2 \Gre^2} \sum_{s_i} \l_{s_i,0}^t (s)
\end{equation}

\begin{equation}\begin{split}\label{M12}
|\M|^2_{1/2} = & \frac{(\a_\dmF)^2}{\L^4} \frac{ m^{2}_\dm s^3 \Big( 1 - \frac{4m_\dm^2}{s} \Big)}{(s-\mre^2)^2+\mre^2 \Gre^2} \sum_{s_i} \l_{s_i,1/2}^t (s) \\
& + \frac{(\b_\dmF \b_5^\dmF)^2}{\L^4} \frac{ m_\dm^2 s^3}{(s-\mim^2)^2+\mim^2 \Gim^2} \sum_{s_i} \l_{s_i,1/2}^a (s)
\end{split}\end{equation}

\begin{equation}\begin{split}\label{M1}
|\M|^2_{1} = & \frac{\a_\dmV^2}{\L^4}\frac{ s^4 \Big( 1 - \frac{4 m_\dm^2}{s} + \frac{6m_\dm^4}{s^2} \Big)}{(s-\mre^2)^2+\mre^2 \Gre^2} \sum_{s_i} \l_{s_i,1}^t (s) \\
& + \frac{\b_\dmV^2}{\L^4} \frac{ s^4 \Big( 1 - \frac{4 m_\dm^2}{s} \Big)}{(s-\mim^2)^2+\mim^2 \Gim^2} \sum_{s_i} \l_{s_i,1}^a (s). 
\end{split}\end{equation}

We parametrize the contribution of the $N_i$ SM initial states with spin $s_i$ for the production of DM of spin $s_f$ through the exchange of a field $j$ by $\l_{s_i,s_f}^j$, which may be functions of the Mandelstam variable $s$, the masses and couplings involved in the processes. They are given in Table \ref{tab:amp}. 

The total production rate of the dark matter candidate with spin $s_f$ from $N_i$ ultra-relativistic ($m_i \ll T, \sqrt{s}$) thermal particles of spin $s_i$, in terms of the Mandelstam variable $s$ is given by \begin{equation}\begin{split}\label{eq:exactrate}
R(T)_{s_f} &= \sum_{s_i} R(T)_{s_i \to s_f} \\ &= \frac{4\pi S_f}{2048 \pi^6} \int_{4m_\dm^2}^\infty ds \sqrt{1-\frac{4m_\dm^2}{s}} \times 
\\ & \times \Big(\sum_{s_i} N_i S_i |\M|_{s_i \to s_f}^2\Big) \int_0^\infty dp_1^i f^i_1 \int_{\frac{s}{4p_{i_1}}}^\infty dp_2^i f_2^i,
\end{split}\end{equation}
where the symmetrization factors $S_{i,f}=1/n_{i,f}!$ account for $n_{i,f}$ identical particles in the initial/final state and $f_{1,2}^i = (e^{-p_{1,2}^i/T} \pm 1)^{-1}$ are the distribution functions of the initial state particles.

The integration over the initial momenta in Eq.\eqref{eq:exactrate} is approximately given by $T \sqrt{s} K_1\Big(\frac{\sqrt{s}}{T}\Big)$, which means that after integrating over $s$ we recognize the light, pole and heavy regimes of the mediator by comparing the \textit{temperature} of the thermal bath with the mediator mass. In Table~\ref{tab:rates} we provide the proportionality constants of the approximate expressions for the rates given in Eqs.~\ref{rate01} and \ref{rate1o2}. 

\newpage

\bibliographystyle{apsrev4-1}
\bibliography{moduli,collabs}

\begin{thebibliography}{61}%
\makeatletter
\providecommand \@ifxundefined [1]{%
 \@ifx{#1\undefined}
}%
\providecommand \@ifnum [1]{%
 \ifnum #1\expandafter \@firstoftwo
 \else \expandafter \@secondoftwo
 \fi
}%
\providecommand \@ifx [1]{%
 \ifx #1\expandafter \@firstoftwo
 \else \expandafter \@secondoftwo
 \fi
}%
\providecommand \natexlab [1]{#1}%
\providecommand \enquote  [1]{``#1''}%
\providecommand \bibnamefont  [1]{#1}%
\providecommand \bibfnamefont [1]{#1}%
\providecommand \citenamefont [1]{#1}%
\providecommand \href@noop [0]{\@secondoftwo}%
\providecommand \href [0]{\begingroup \@sanitize@url \@href}%
\providecommand \@href[1]{\@@startlink{#1}\@@href}%
\providecommand \@@href[1]{\endgroup#1\@@endlink}%
\providecommand \@sanitize@url [0]{\catcode `\\12\catcode `\$12\catcode
  `\&12\catcode `\#12\catcode `\^12\catcode `\_12\catcode `\%12\relax}%
\providecommand \@@startlink[1]{}%
\providecommand \@@endlink[0]{}%
\providecommand \url  [0]{\begingroup\@sanitize@url \@url }%
\providecommand \@url [1]{\endgroup\@href {#1}{\urlprefix }}%
\providecommand \urlprefix  [0]{URL }%
\providecommand \Eprint [0]{\href }%
\providecommand \doibase [0]{http://dx.doi.org/}%
\providecommand \selectlanguage [0]{\@gobble}%
\providecommand \bibinfo  [0]{\@secondoftwo}%
\providecommand \bibfield  [0]{\@secondoftwo}%
\providecommand \translation [1]{[#1]}%
\providecommand \BibitemOpen [0]{}%
\providecommand \bibitemStop [0]{}%
\providecommand \bibitemNoStop [0]{.\EOS\space}%
\providecommand \EOS [0]{\spacefactor3000\relax}%
\providecommand \BibitemShut  [1]{\csname bibitem#1\endcsname}%
\let\auto@bib@innerbib\@empty
\bibitem [{\citenamefont {{\relax Planck
  Collaboration}}(2018)}]{planck_collaboration_planck_2018}%
  \BibitemOpen
  \bibfield  {author} {\bibinfo {author} {\bibnamefont {{\relax Planck
  Collaboration}}},\ }\href@noop {} {\  (\bibinfo {year} {2018})},\ \Eprint
  {http://arxiv.org/abs/1807.06209} {arXiv:1807.06209 [astro-ph.CO]}
  \BibitemShut {NoStop}%
\bibitem [{\citenamefont {{\relax XENON100
  Collaboration}}(2012)}]{xenon100_collaboration_dark_2012_ok}%
  \BibitemOpen
  \bibfield  {author} {\bibinfo {author} {\bibnamefont {{\relax XENON100
  Collaboration}}},\ }\href {\doibase 10.1103/PhysRevLett.109.181301}
  {\bibfield  {journal} {\bibinfo  {journal} {Physical Review Letters}\
  }\textbf {\bibinfo {volume} {109}} (\bibinfo {year} {2012}),\
  10.1103/PhysRevLett.109.181301},\ \bibinfo {note} {arXiv:
  1207.5988}\BibitemShut {NoStop}%
\bibitem [{\citenamefont {{\relax LUX
  Collaboration}}(2017)}]{akerib_results_2017_ok}%
  \BibitemOpen
  \bibfield  {author} {\bibinfo {author} {\bibnamefont {{\relax LUX
  Collaboration}}},\ }\href {\doibase 10.1103/PhysRevLett.118.021303}
  {\bibfield  {journal} {\bibinfo  {journal} {Physical Review Letters}\
  }\textbf {\bibinfo {volume} {118}} (\bibinfo {year} {2017}),\
  10.1103/PhysRevLett.118.021303},\ \bibinfo {note} {arXiv:
  1608.07648}\BibitemShut {NoStop}%
\bibitem [{\citenamefont {{\relax PANDAX-II
  Collaboration}}(2017)}]{fu_spin-dependent_2017_ok}%
  \BibitemOpen
  \bibfield  {author} {\bibinfo {author} {\bibnamefont {{\relax PANDAX-II
  Collaboration}}},\ }\href {\doibase 10.1103/PhysRevLett.118.071301}
  {\bibfield  {journal} {\bibinfo  {journal} {Physical Review Letters}\
  }\textbf {\bibinfo {volume} {118}} (\bibinfo {year} {2017}),\
  10.1103/PhysRevLett.118.071301},\ \bibinfo {note} {arXiv:
  1611.06553}\BibitemShut {NoStop}%
\bibitem [{\citenamefont {Casas}\ \emph {et~al.}(2017)\citenamefont {Casas},
  \citenamefont {Cerde{\~n}o}, \citenamefont {Moreno},\ and\ \citenamefont
  {Quilis}}]{casas_reopening_2017}%
  \BibitemOpen
  \bibfield  {author} {\bibinfo {author} {\bibfnamefont {J.~A.}\ \bibnamefont
  {Casas}}, \bibinfo {author} {\bibfnamefont {D.~G.}\ \bibnamefont
  {Cerde{\~n}o}}, \bibinfo {author} {\bibfnamefont {J.~M.}\ \bibnamefont
  {Moreno}}, \ and\ \bibinfo {author} {\bibfnamefont {J.}~\bibnamefont
  {Quilis}},\ }\href {\doibase 10.1007/JHEP05(2017)036} {\bibfield  {journal}
  {\bibinfo  {journal} {Journal of High Energy Physics}\ }\textbf {\bibinfo
  {volume} {2017}} (\bibinfo {year} {2017}),\ 10.1007/JHEP05(2017)036},\
  \bibinfo {note} {arXiv: 1701.08134}\BibitemShut {NoStop}%
\bibitem [{\citenamefont {Djouadi}\ \emph {et~al.}(2012)\citenamefont
  {Djouadi}, \citenamefont {Lebedev}, \citenamefont {Mambrini},\ and\
  \citenamefont {Quevillon}}]{djouadi_implications_2012}%
  \BibitemOpen
  \bibfield  {author} {\bibinfo {author} {\bibfnamefont {A.}~\bibnamefont
  {Djouadi}}, \bibinfo {author} {\bibfnamefont {O.}~\bibnamefont {Lebedev}},
  \bibinfo {author} {\bibfnamefont {Y.}~\bibnamefont {Mambrini}}, \ and\
  \bibinfo {author} {\bibfnamefont {J.}~\bibnamefont {Quevillon}},\ }\href
  {\doibase 10.1016/j.physletb.2012.01.062} {\bibfield  {journal} {\bibinfo
  {journal} {Physics Letters B}\ }\textbf {\bibinfo {volume} {709}},\ \bibinfo
  {pages} {65} (\bibinfo {year} {2012})},\ \bibinfo {note} {arXiv:
  1112.3299}\BibitemShut {NoStop}%
\bibitem [{\citenamefont {Djouadi}\ \emph {et~al.}(2013)\citenamefont
  {Djouadi}, \citenamefont {Falkowski}, \citenamefont {Mambrini},\ and\
  \citenamefont {Quevillon}}]{djouadi_direct_2013}%
  \BibitemOpen
  \bibfield  {author} {\bibinfo {author} {\bibfnamefont {A.}~\bibnamefont
  {Djouadi}}, \bibinfo {author} {\bibfnamefont {A.}~\bibnamefont {Falkowski}},
  \bibinfo {author} {\bibfnamefont {Y.}~\bibnamefont {Mambrini}}, \ and\
  \bibinfo {author} {\bibfnamefont {J.}~\bibnamefont {Quevillon}},\ }\href
  {\doibase 10.1140/epjc/s10052-013-2455-1} {\bibfield  {journal} {\bibinfo
  {journal} {The European Physical Journal C}\ }\textbf {\bibinfo {volume}
  {73}} (\bibinfo {year} {2013}),\ 10.1140/epjc/s10052-013-2455-1},\ \bibinfo
  {note} {arXiv: 1205.3169}\BibitemShut {NoStop}%
\bibitem [{\citenamefont {Lebedev}\ \emph {et~al.}(2012)\citenamefont
  {Lebedev}, \citenamefont {Lee},\ and\ \citenamefont
  {Mambrini}}]{lebedev_vector_2012}%
  \BibitemOpen
  \bibfield  {author} {\bibinfo {author} {\bibfnamefont {O.}~\bibnamefont
  {Lebedev}}, \bibinfo {author} {\bibfnamefont {H.~M.}\ \bibnamefont {Lee}}, \
  and\ \bibinfo {author} {\bibfnamefont {Y.}~\bibnamefont {Mambrini}},\ }\href
  {\doibase 10.1016/j.physletb.2012.01.029} {\bibfield  {journal} {\bibinfo
  {journal} {Physics Letters B}\ }\textbf {\bibinfo {volume} {707}},\ \bibinfo
  {pages} {570} (\bibinfo {year} {2012})},\ \bibinfo {note} {arXiv:
  1111.4482}\BibitemShut {NoStop}%
\bibitem [{\citenamefont {Mambrini}(2011)}]{mambrini_higgs_2011}%
  \BibitemOpen
  \bibfield  {author} {\bibinfo {author} {\bibfnamefont {Y.}~\bibnamefont
  {Mambrini}},\ }\href {\doibase 10.1103/PhysRevD.84.115017} {\bibfield
  {journal} {\bibinfo  {journal} {Physical Review D}\ }\textbf {\bibinfo
  {volume} {84}} (\bibinfo {year} {2011}),\ 10.1103/PhysRevD.84.115017},\
  \bibinfo {note} {arXiv: 1108.0671}\BibitemShut {NoStop}%
\bibitem [{\citenamefont {Ellis}\ \emph {et~al.}(2018)\citenamefont {Ellis},
  \citenamefont {Fowlie}, \citenamefont {Marzola},\ and\ \citenamefont
  {Raidal}}]{ellis_statistical_2018}%
  \BibitemOpen
  \bibfield  {author} {\bibinfo {author} {\bibfnamefont {J.}~\bibnamefont
  {Ellis}}, \bibinfo {author} {\bibfnamefont {A.}~\bibnamefont {Fowlie}},
  \bibinfo {author} {\bibfnamefont {L.}~\bibnamefont {Marzola}}, \ and\
  \bibinfo {author} {\bibfnamefont {M.}~\bibnamefont {Raidal}},\ }\href
  {\doibase 10.1103/PhysRevD.97.115014} {\bibfield  {journal} {\bibinfo
  {journal} {Physical Review D}\ }\textbf {\bibinfo {volume} {97}} (\bibinfo
  {year} {2018}),\ 10.1103/PhysRevD.97.115014},\ \bibinfo {note} {arXiv:
  1711.09912}\BibitemShut {NoStop}%
\bibitem [{\citenamefont {Arcadi}\ \emph {et~al.}(2015)\citenamefont {Arcadi},
  \citenamefont {Mambrini},\ and\ \citenamefont
  {Richard}}]{arcadi_z-portal_2015}%
  \BibitemOpen
  \bibfield  {author} {\bibinfo {author} {\bibfnamefont {G.}~\bibnamefont
  {Arcadi}}, \bibinfo {author} {\bibfnamefont {Y.}~\bibnamefont {Mambrini}}, \
  and\ \bibinfo {author} {\bibfnamefont {F.}~\bibnamefont {Richard}},\ }\href
  {\doibase 10.1088/1475-7516/2015/03/018} {\bibfield  {journal} {\bibinfo
  {journal} {Journal of Cosmology and Astroparticle Physics}\ }\textbf
  {\bibinfo {volume} {2015}},\ \bibinfo {pages} {018} (\bibinfo {year}
  {2015})},\ \bibinfo {note} {arXiv: 1411.2985}\BibitemShut {NoStop}%
\bibitem [{\citenamefont {Kearney}\ \emph {et~al.}(2017)\citenamefont
  {Kearney}, \citenamefont {Orlofsky},\ and\ \citenamefont
  {Pierce}}]{kearney_$z$_2017}%
  \BibitemOpen
  \bibfield  {author} {\bibinfo {author} {\bibfnamefont {J.}~\bibnamefont
  {Kearney}}, \bibinfo {author} {\bibfnamefont {N.}~\bibnamefont {Orlofsky}}, \
  and\ \bibinfo {author} {\bibfnamefont {A.}~\bibnamefont {Pierce}},\ }\href
  {\doibase 10.1103/PhysRevD.95.035020} {\bibfield  {journal} {\bibinfo
  {journal} {Physical Review D}\ }\textbf {\bibinfo {volume} {95}} (\bibinfo
  {year} {2017}),\ 10.1103/PhysRevD.95.035020},\ \bibinfo {note} {arXiv:
  1611.05048}\BibitemShut {NoStop}%
\bibitem [{\citenamefont {Escudero}\ \emph {et~al.}(2016)\citenamefont
  {Escudero}, \citenamefont {Berlin}, \citenamefont {Hooper},\ and\
  \citenamefont {Lin}}]{escudero_toward_2016}%
  \BibitemOpen
  \bibfield  {author} {\bibinfo {author} {\bibfnamefont {M.}~\bibnamefont
  {Escudero}}, \bibinfo {author} {\bibfnamefont {A.}~\bibnamefont {Berlin}},
  \bibinfo {author} {\bibfnamefont {D.}~\bibnamefont {Hooper}}, \ and\ \bibinfo
  {author} {\bibfnamefont {M.-X.}\ \bibnamefont {Lin}},\ }\href {\doibase
  10.1088/1475-7516/2016/12/029} {\bibfield  {journal} {\bibinfo  {journal}
  {Journal of Cosmology and Astroparticle Physics}\ }\textbf {\bibinfo {volume}
  {2016}},\ \bibinfo {pages} {029} (\bibinfo {year} {2016})},\ \bibinfo {note}
  {arXiv: 1609.09079}\BibitemShut {NoStop}%
\bibitem [{\citenamefont {Alves}\ \emph {et~al.}(2014)\citenamefont {Alves},
  \citenamefont {Profumo},\ and\ \citenamefont {Queiroz}}]{alves_dark_2014}%
  \BibitemOpen
  \bibfield  {author} {\bibinfo {author} {\bibfnamefont {A.}~\bibnamefont
  {Alves}}, \bibinfo {author} {\bibfnamefont {S.}~\bibnamefont {Profumo}}, \
  and\ \bibinfo {author} {\bibfnamefont {F.~S.}\ \bibnamefont {Queiroz}},\
  }\href {\doibase 10.1007/JHEP04(2014)063} {\bibfield  {journal} {\bibinfo
  {journal} {Journal of High Energy Physics}\ }\textbf {\bibinfo {volume}
  {2014}} (\bibinfo {year} {2014}),\ 10.1007/JHEP04(2014)063},\ \bibinfo {note}
  {arXiv: 1312.5281}\BibitemShut {NoStop}%
\bibitem [{\citenamefont {Lebedev}\ and\ \citenamefont
  {Mambrini}(2014)}]{lebedev_axial_2014}%
  \BibitemOpen
  \bibfield  {author} {\bibinfo {author} {\bibfnamefont {O.}~\bibnamefont
  {Lebedev}}\ and\ \bibinfo {author} {\bibfnamefont {Y.}~\bibnamefont
  {Mambrini}},\ }\href {\doibase 10.1016/j.physletb.2014.05.025} {\bibfield
  {journal} {\bibinfo  {journal} {Physics Letters B}\ }\textbf {\bibinfo
  {volume} {734}},\ \bibinfo {pages} {350} (\bibinfo {year} {2014})},\ \bibinfo
  {note} {arXiv: 1403.4837}\BibitemShut {NoStop}%
\bibitem [{\citenamefont {Arcadi}\ \emph {et~al.}(2014)\citenamefont {Arcadi},
  \citenamefont {Mambrini}, \citenamefont {Tytgat},\ and\ \citenamefont
  {Zaldivar}}]{arcadi_invisible_2014}%
  \BibitemOpen
  \bibfield  {author} {\bibinfo {author} {\bibfnamefont {G.}~\bibnamefont
  {Arcadi}}, \bibinfo {author} {\bibfnamefont {Y.}~\bibnamefont {Mambrini}},
  \bibinfo {author} {\bibfnamefont {M.~H.~G.}\ \bibnamefont {Tytgat}}, \ and\
  \bibinfo {author} {\bibfnamefont {B.}~\bibnamefont {Zaldivar}},\ }\href
  {\doibase 10.1007/JHEP03(2014)134} {\bibfield  {journal} {\bibinfo  {journal}
  {Journal of High Energy Physics}\ }\textbf {\bibinfo {volume} {2014}}
  (\bibinfo {year} {2014}),\ 10.1007/JHEP03(2014)134},\ \bibinfo {note} {arXiv:
  1401.0221}\BibitemShut {NoStop}%
\bibitem [{\citenamefont {Arcadi}\ \emph {et~al.}(2018)\citenamefont {Arcadi},
  \citenamefont {Dutra}, \citenamefont {Ghosh}, \citenamefont {Lindner},
  \citenamefont {Mambrini}, \citenamefont {Pierre}, \citenamefont {Profumo},\
  and\ \citenamefont {Queiroz}}]{arcadi_waning_2018}%
  \BibitemOpen
  \bibfield  {author} {\bibinfo {author} {\bibfnamefont {G.}~\bibnamefont
  {Arcadi}}, \bibinfo {author} {\bibfnamefont {M.}~\bibnamefont {Dutra}},
  \bibinfo {author} {\bibfnamefont {P.}~\bibnamefont {Ghosh}}, \bibinfo
  {author} {\bibfnamefont {M.}~\bibnamefont {Lindner}}, \bibinfo {author}
  {\bibfnamefont {Y.}~\bibnamefont {Mambrini}}, \bibinfo {author}
  {\bibfnamefont {M.}~\bibnamefont {Pierre}}, \bibinfo {author} {\bibfnamefont
  {S.}~\bibnamefont {Profumo}}, \ and\ \bibinfo {author} {\bibfnamefont
  {F.~S.}\ \bibnamefont {Queiroz}},\ }\href {\doibase
  10.1140/epjc/s10052-018-5662-y} {\bibfield  {journal} {\bibinfo  {journal}
  {The European Physical Journal C}\ }\textbf {\bibinfo {volume} {78}}
  (\bibinfo {year} {2018}),\ 10.1140/epjc/s10052-018-5662-y},\ \bibinfo {note}
  {arXiv: 1703.07364}\BibitemShut {NoStop}%
\bibitem [{\citenamefont {Hall}\ \emph {et~al.}(2010)\citenamefont {Hall},
  \citenamefont {Jedamzik}, \citenamefont {March-Russell},\ and\ \citenamefont
  {West}}]{hall_freeze-production_2010}%
  \BibitemOpen
  \bibfield  {author} {\bibinfo {author} {\bibfnamefont {L.~J.}\ \bibnamefont
  {Hall}}, \bibinfo {author} {\bibfnamefont {K.}~\bibnamefont {Jedamzik}},
  \bibinfo {author} {\bibfnamefont {J.}~\bibnamefont {March-Russell}}, \ and\
  \bibinfo {author} {\bibfnamefont {S.~M.}\ \bibnamefont {West}},\ }\href
  {\doibase 10.1007/JHEP03(2010)080} {\bibfield  {journal} {\bibinfo  {journal}
  {Journal of High Energy Physics}\ }\textbf {\bibinfo {volume} {2010}}
  (\bibinfo {year} {2010}),\ 10.1007/JHEP03(2010)080},\ \bibinfo {note} {arXiv:
  0911.1120}\BibitemShut {NoStop}%
\bibitem [{\citenamefont {Chu}\ \emph {et~al.}(2012)\citenamefont {Chu},
  \citenamefont {Hambye},\ and\ \citenamefont {Tytgat}}]{chu_four_2012}%
  \BibitemOpen
  \bibfield  {author} {\bibinfo {author} {\bibfnamefont {X.}~\bibnamefont
  {Chu}}, \bibinfo {author} {\bibfnamefont {T.}~\bibnamefont {Hambye}}, \ and\
  \bibinfo {author} {\bibfnamefont {M.~H.~G.}\ \bibnamefont {Tytgat}},\ }\href
  {\doibase 10.1088/1475-7516/2012/05/034} {\bibfield  {journal} {\bibinfo
  {journal} {Journal of Cosmology and Astroparticle Physics}\ }\textbf
  {\bibinfo {volume} {2012}},\ \bibinfo {pages} {034} (\bibinfo {year}
  {2012})},\ \bibinfo {note} {arXiv: 1112.0493}\BibitemShut {NoStop}%
\bibitem [{\citenamefont {Chu}\ \emph {et~al.}(2014)\citenamefont {Chu},
  \citenamefont {Mambrini}, \citenamefont {Quevillon},\ and\ \citenamefont
  {Zaldivar}}]{chu_thermal_2014}%
  \BibitemOpen
  \bibfield  {author} {\bibinfo {author} {\bibfnamefont {X.}~\bibnamefont
  {Chu}}, \bibinfo {author} {\bibfnamefont {Y.}~\bibnamefont {Mambrini}},
  \bibinfo {author} {\bibfnamefont {J.}~\bibnamefont {Quevillon}}, \ and\
  \bibinfo {author} {\bibfnamefont {B.}~\bibnamefont {Zaldivar}},\ }\href
  {\doibase 10.1088/1475-7516/2014/01/034} {\bibfield  {journal} {\bibinfo
  {journal} {Journal of Cosmology and Astroparticle Physics}\ }\textbf
  {\bibinfo {volume} {2014}},\ \bibinfo {pages} {034} (\bibinfo {year}
  {2014})},\ \bibinfo {note} {arXiv: 1306.4677}\BibitemShut {NoStop}%
\bibitem [{\citenamefont {Bernal}\ \emph {et~al.}(2017)\citenamefont {Bernal},
  \citenamefont {Heikinheimo}, \citenamefont {Tenkanen}, \citenamefont
  {Tuominen},\ and\ \citenamefont {Vaskonen}}]{bernal_dawn_2017}%
  \BibitemOpen
  \bibfield  {author} {\bibinfo {author} {\bibfnamefont {N.}~\bibnamefont
  {Bernal}}, \bibinfo {author} {\bibfnamefont {M.}~\bibnamefont {Heikinheimo}},
  \bibinfo {author} {\bibfnamefont {T.}~\bibnamefont {Tenkanen}}, \bibinfo
  {author} {\bibfnamefont {K.}~\bibnamefont {Tuominen}}, \ and\ \bibinfo
  {author} {\bibfnamefont {V.}~\bibnamefont {Vaskonen}},\ }\href {\doibase
  10.1142/S0217751X1730023X} {\bibfield  {journal} {\bibinfo  {journal}
  {International Journal of Modern Physics A}\ }\textbf {\bibinfo {volume}
  {32}},\ \bibinfo {pages} {1730023} (\bibinfo {year} {2017})},\ \bibinfo
  {note} {arXiv: 1706.07442}\BibitemShut {NoStop}%
\bibitem [{\citenamefont {Mambrini}\ \emph {et~al.}(2013)\citenamefont
  {Mambrini}, \citenamefont {Olive}, \citenamefont {Quevillon},\ and\
  \citenamefont {Zaldivar}}]{mambrini_gauge_2013}%
  \BibitemOpen
  \bibfield  {author} {\bibinfo {author} {\bibfnamefont {Y.}~\bibnamefont
  {Mambrini}}, \bibinfo {author} {\bibfnamefont {K.~A.}\ \bibnamefont {Olive}},
  \bibinfo {author} {\bibfnamefont {J.}~\bibnamefont {Quevillon}}, \ and\
  \bibinfo {author} {\bibfnamefont {B.}~\bibnamefont {Zaldivar}},\ }\href
  {\doibase 10.1103/PhysRevLett.110.241306} {\bibfield  {journal} {\bibinfo
  {journal} {Physical Review Letters}\ }\textbf {\bibinfo {volume} {110}}
  (\bibinfo {year} {2013}),\ 10.1103/PhysRevLett.110.241306},\ \bibinfo {note}
  {arXiv: 1302.4438}\BibitemShut {NoStop}%
\bibitem [{\citenamefont {Mambrini}\ \emph {et~al.}(2015)\citenamefont
  {Mambrini}, \citenamefont {Nagata}, \citenamefont {Olive}, \citenamefont
  {Quevillon},\ and\ \citenamefont {Zheng}}]{mambrini_dark_2015}%
  \BibitemOpen
  \bibfield  {author} {\bibinfo {author} {\bibfnamefont {Y.}~\bibnamefont
  {Mambrini}}, \bibinfo {author} {\bibfnamefont {N.}~\bibnamefont {Nagata}},
  \bibinfo {author} {\bibfnamefont {K.~A.}\ \bibnamefont {Olive}}, \bibinfo
  {author} {\bibfnamefont {J.}~\bibnamefont {Quevillon}}, \ and\ \bibinfo
  {author} {\bibfnamefont {J.}~\bibnamefont {Zheng}},\ }\href {\doibase
  10.1103/PhysRevD.91.095010} {\  (\bibinfo {year} {2015}),\
  10.1103/PhysRevD.91.095010}\BibitemShut {NoStop}%
\bibitem [{\citenamefont {Mambrini}\ \emph {et~al.}(2016)\citenamefont
  {Mambrini}, \citenamefont {Nagata}, \citenamefont {Olive},\ and\
  \citenamefont {Zheng}}]{mambrini_vacuum_2016}%
  \BibitemOpen
  \bibfield  {author} {\bibinfo {author} {\bibfnamefont {Y.}~\bibnamefont
  {Mambrini}}, \bibinfo {author} {\bibfnamefont {N.}~\bibnamefont {Nagata}},
  \bibinfo {author} {\bibfnamefont {K.~A.}\ \bibnamefont {Olive}}, \ and\
  \bibinfo {author} {\bibfnamefont {J.}~\bibnamefont {Zheng}},\ }\href
  {\doibase 10.1103/PhysRevD.93.111703} {\  (\bibinfo {year} {2016}),\
  10.1103/PhysRevD.93.111703}\BibitemShut {NoStop}%
\bibitem [{\citenamefont {Bhattacharyya}\ \emph {et~al.}(2018)\citenamefont
  {Bhattacharyya}, \citenamefont {Dutra}, \citenamefont {Mambrini},\ and\
  \citenamefont {Pierre}}]{bhattacharyya_freezing-dark_2018}%
  \BibitemOpen
  \bibfield  {author} {\bibinfo {author} {\bibfnamefont {G.}~\bibnamefont
  {Bhattacharyya}}, \bibinfo {author} {\bibfnamefont {M.}~\bibnamefont
  {Dutra}}, \bibinfo {author} {\bibfnamefont {Y.}~\bibnamefont {Mambrini}}, \
  and\ \bibinfo {author} {\bibfnamefont {M.}~\bibnamefont {Pierre}},\ }\href
  {\doibase 10.1103/PhysRevD.98.035038} {\bibfield  {journal} {\bibinfo
  {journal} {Physical Review D}\ }\textbf {\bibinfo {volume} {98}} (\bibinfo
  {year} {2018}),\ 10.1103/PhysRevD.98.035038},\ \bibinfo {note} {arXiv:
  1806.00016}\BibitemShut {NoStop}%
\bibitem [{\citenamefont {Bernal}\ \emph {et~al.}(2018)\citenamefont {Bernal},
  \citenamefont {Dutra}, \citenamefont {Mambrini}, \citenamefont {Olive},
  \citenamefont {Peloso},\ and\ \citenamefont {Pierre}}]{bernal_spin-2_2018}%
  \BibitemOpen
  \bibfield  {author} {\bibinfo {author} {\bibfnamefont {N.}~\bibnamefont
  {Bernal}}, \bibinfo {author} {\bibfnamefont {M.}~\bibnamefont {Dutra}},
  \bibinfo {author} {\bibfnamefont {Y.}~\bibnamefont {Mambrini}}, \bibinfo
  {author} {\bibfnamefont {K.~A.}\ \bibnamefont {Olive}}, \bibinfo {author}
  {\bibfnamefont {M.}~\bibnamefont {Peloso}}, \ and\ \bibinfo {author}
  {\bibfnamefont {M.}~\bibnamefont {Pierre}},\ }\href {\doibase
  10.1103/PhysRevD.97.115020} {\bibfield  {journal} {\bibinfo  {journal}
  {Physical Review D}\ }\textbf {\bibinfo {volume} {97}} (\bibinfo {year}
  {2018}),\ 10.1103/PhysRevD.97.115020},\ \bibinfo {note} {arXiv:
  1803.01866}\BibitemShut {NoStop}%
\bibitem [{\citenamefont {Benakli}\ \emph {et~al.}(2017)\citenamefont
  {Benakli}, \citenamefont {Chen}, \citenamefont {Dudas},\ and\ \citenamefont
  {Mambrini}}]{benakli_minimal_2017}%
  \BibitemOpen
  \bibfield  {author} {\bibinfo {author} {\bibfnamefont {K.}~\bibnamefont
  {Benakli}}, \bibinfo {author} {\bibfnamefont {Y.}~\bibnamefont {Chen}},
  \bibinfo {author} {\bibfnamefont {E.}~\bibnamefont {Dudas}}, \ and\ \bibinfo
  {author} {\bibfnamefont {Y.}~\bibnamefont {Mambrini}},\ }\href {\doibase
  10.1103/PhysRevD.95.095002} {\bibfield  {journal} {\bibinfo  {journal}
  {Physical Review D}\ }\textbf {\bibinfo {volume} {95}} (\bibinfo {year}
  {2017}),\ 10.1103/PhysRevD.95.095002},\ \bibinfo {note} {arXiv:
  1701.06574}\BibitemShut {NoStop}%
\bibitem [{\citenamefont {Dudas}\ \emph
  {et~al.}(2017{\natexlab{a}})\citenamefont {Dudas}, \citenamefont {Mambrini},\
  and\ \citenamefont {Olive}}]{dudas_case_2017}%
  \BibitemOpen
  \bibfield  {author} {\bibinfo {author} {\bibfnamefont {E.}~\bibnamefont
  {Dudas}}, \bibinfo {author} {\bibfnamefont {Y.}~\bibnamefont {Mambrini}}, \
  and\ \bibinfo {author} {\bibfnamefont {K.}~\bibnamefont {Olive}},\ }\href
  {\doibase 10.1103/PhysRevLett.119.051801} {\bibfield  {journal} {\bibinfo
  {journal} {Physical Review Letters}\ }\textbf {\bibinfo {volume} {119}}
  (\bibinfo {year} {2017}{\natexlab{a}}),\ 10.1103/PhysRevLett.119.051801},\
  \bibinfo {note} {arXiv: 1704.03008}\BibitemShut {NoStop}%
\bibitem [{\citenamefont {Dudas}\ \emph
  {et~al.}(2017{\natexlab{b}})\citenamefont {Dudas}, \citenamefont
  {Gherghetta}, \citenamefont {Mambrini},\ and\ \citenamefont
  {Olive}}]{dudas_inflation_2017}%
  \BibitemOpen
  \bibfield  {author} {\bibinfo {author} {\bibfnamefont {E.}~\bibnamefont
  {Dudas}}, \bibinfo {author} {\bibfnamefont {T.}~\bibnamefont {Gherghetta}},
  \bibinfo {author} {\bibfnamefont {Y.}~\bibnamefont {Mambrini}}, \ and\
  \bibinfo {author} {\bibfnamefont {K.~A.}\ \bibnamefont {Olive}},\ }\href
  {\doibase 10.1103/PhysRevD.96.115032} {\  (\bibinfo {year}
  {2017}{\natexlab{b}}),\ 10.1103/PhysRevD.96.115032}\BibitemShut {NoStop}%
\bibitem [{\citenamefont {Dudas}\ \emph {et~al.}(2018)\citenamefont {Dudas},
  \citenamefont {Gherghetta}, \citenamefont {Kaneta}, \citenamefont
  {Mambrini},\ and\ \citenamefont {Olive}}]{dudas_gravitino_2018}%
  \BibitemOpen
  \bibfield  {author} {\bibinfo {author} {\bibfnamefont {E.}~\bibnamefont
  {Dudas}}, \bibinfo {author} {\bibfnamefont {T.}~\bibnamefont {Gherghetta}},
  \bibinfo {author} {\bibfnamefont {K.}~\bibnamefont {Kaneta}}, \bibinfo
  {author} {\bibfnamefont {Y.}~\bibnamefont {Mambrini}}, \ and\ \bibinfo
  {author} {\bibfnamefont {K.~A.}\ \bibnamefont {Olive}},\ }\href {\doibase
  10.1103/PhysRevD.98.015030} {\  (\bibinfo {year} {2018}),\
  10.1103/PhysRevD.98.015030}\BibitemShut {NoStop}%
\bibitem [{\citenamefont {Garcia}\ \emph {et~al.}(2017)\citenamefont {Garcia},
  \citenamefont {Mambrini}, \citenamefont {Olive},\ and\ \citenamefont
  {Peloso}}]{garcia_enhancement_2017}%
  \BibitemOpen
  \bibfield  {author} {\bibinfo {author} {\bibfnamefont {M.~A.~G.}\
  \bibnamefont {Garcia}}, \bibinfo {author} {\bibfnamefont {Y.}~\bibnamefont
  {Mambrini}}, \bibinfo {author} {\bibfnamefont {K.~A.}\ \bibnamefont {Olive}},
  \ and\ \bibinfo {author} {\bibfnamefont {M.}~\bibnamefont {Peloso}},\ }\href
  {\doibase 10.1103/PhysRevD.96.103510} {\bibfield  {journal} {\bibinfo
  {journal} {Physical Review D}\ }\textbf {\bibinfo {volume} {96}} (\bibinfo
  {year} {2017}),\ 10.1103/PhysRevD.96.103510},\ \bibinfo {note} {arXiv:
  1709.01549}\BibitemShut {NoStop}%
\bibitem [{\citenamefont {Garcia}\ and\ \citenamefont
  {Amin}(2018)}]{garcia_pre-thermalization_2018}%
  \BibitemOpen
  \bibfield  {author} {\bibinfo {author} {\bibfnamefont {M.~A.~G.}\
  \bibnamefont {Garcia}}\ and\ \bibinfo {author} {\bibfnamefont {M.~A.}\
  \bibnamefont {Amin}},\ }\href {https://arxiv.org/abs/1806.01865} {\
  (\bibinfo {year} {2018})}\BibitemShut {NoStop}%
\bibitem [{\citenamefont {Taylor}\ and\ \citenamefont
  {Veneziano}(1988)}]{taylor_dilaton_1988}%
  \BibitemOpen
  \bibfield  {author} {\bibinfo {author} {\bibfnamefont {T.~R.}\ \bibnamefont
  {Taylor}}\ and\ \bibinfo {author} {\bibfnamefont {G.}~\bibnamefont
  {Veneziano}},\ }\href {\doibase 10.1016/0370-2693(88)91290-7} {\bibfield
  {journal} {\bibinfo  {journal} {Physics Letters B}\ }\textbf {\bibinfo
  {volume} {213}},\ \bibinfo {pages} {450} (\bibinfo {year}
  {1988})}\BibitemShut {NoStop}%
\bibitem [{\citenamefont {Damour}\ and\ \citenamefont
  {Polyakov}(1994)}]{damour_string_1994}%
  \BibitemOpen
  \bibfield  {author} {\bibinfo {author} {\bibfnamefont {T.}~\bibnamefont
  {Damour}}\ and\ \bibinfo {author} {\bibfnamefont {A.~M.}\ \bibnamefont
  {Polyakov}},\ }\href {\doibase 10.1016/0550-3213(94)90143-0} {\bibfield
  {journal} {\bibinfo  {journal} {Nuclear Physics B}\ }\textbf {\bibinfo
  {volume} {423}},\ \bibinfo {pages} {532} (\bibinfo {year} {1994})},\ \bibinfo
  {note} {arXiv: hep-th/9401069}\BibitemShut {NoStop}%
\bibitem [{\citenamefont {Binetruy}\ and\ \citenamefont
  {Dudas}(1994)}]{binetruy_nambu_1994}%
  \BibitemOpen
  \bibfield  {author} {\bibinfo {author} {\bibfnamefont {P.}~\bibnamefont
  {Binetruy}}\ and\ \bibinfo {author} {\bibfnamefont {E.}~\bibnamefont
  {Dudas}},\ }\href {\doibase 10.1016/0370-2693(94)91338-2} {\bibfield
  {journal} {\bibinfo  {journal} {Physics Letters B}\ }\textbf {\bibinfo
  {volume} {338}},\ \bibinfo {pages} {23} (\bibinfo {year} {1994})},\ \bibinfo
  {note} {arXiv: hep-ph/9405429}\BibitemShut {NoStop}%
\bibitem [{\citenamefont {Binetruy}\ and\ \citenamefont
  {Dudas}(1995{\natexlab{a}})}]{Binetruy:1994bn}%
  \BibitemOpen
  \bibfield  {author} {\bibinfo {author} {\bibfnamefont {P.}~\bibnamefont
  {Binetruy}}\ and\ \bibinfo {author} {\bibfnamefont {E.}~\bibnamefont
  {Dudas}},\ }\href {\doibase 10.1016/S0550-3213(95)00069-0} {\bibfield
  {journal} {\bibinfo  {journal} {Nucl. Phys.}\ }\textbf {\bibinfo {volume}
  {B442}},\ \bibinfo {pages} {21} (\bibinfo {year} {1995}{\natexlab{a}})},\
  \Eprint {http://arxiv.org/abs/hep-ph/9411413} {arXiv:hep-ph/9411413 [hep-ph]}
  \BibitemShut {NoStop}%
\bibitem [{\citenamefont {Binetruy}\ and\ \citenamefont
  {Dudas}(1995{\natexlab{b}})}]{Binetruy:1995nt}%
  \BibitemOpen
  \bibfield  {author} {\bibinfo {author} {\bibfnamefont {P.}~\bibnamefont
  {Binetruy}}\ and\ \bibinfo {author} {\bibfnamefont {E.}~\bibnamefont
  {Dudas}},\ }\href {\doibase 10.1016/0550-3213(95)00345-S} {\bibfield
  {journal} {\bibinfo  {journal} {Nucl. Phys.}\ }\textbf {\bibinfo {volume}
  {B451}},\ \bibinfo {pages} {31} (\bibinfo {year} {1995}{\natexlab{b}})},\
  \Eprint {http://arxiv.org/abs/hep-ph/9505295} {arXiv:hep-ph/9505295 [hep-ph]}
  \BibitemShut {NoStop}%
\bibitem [{\citenamefont {Kounnas}\ \emph {et~al.}(1994)\citenamefont
  {Kounnas}, \citenamefont {Pavel},\ and\ \citenamefont
  {Zwirner}}]{kounnas_towards_1994}%
  \BibitemOpen
  \bibfield  {author} {\bibinfo {author} {\bibfnamefont {C.}~\bibnamefont
  {Kounnas}}, \bibinfo {author} {\bibfnamefont {I.}~\bibnamefont {Pavel}}, \
  and\ \bibinfo {author} {\bibfnamefont {F.}~\bibnamefont {Zwirner}},\ }\href
  {\doibase 10.1016/0370-2693(94)90371-9} {\bibfield  {journal} {\bibinfo
  {journal} {Physics Letters B}\ }\textbf {\bibinfo {volume} {335}},\ \bibinfo
  {pages} {403} (\bibinfo {year} {1994})},\ \bibinfo {note} {arXiv:
  hep-ph/9406256}\BibitemShut {NoStop}%
\bibitem [{\citenamefont {Dimopoulos}\ and\ \citenamefont
  {Giudice}(1996)}]{dimopoulos_macroscopic_1996}%
  \BibitemOpen
  \bibfield  {author} {\bibinfo {author} {\bibfnamefont {S.}~\bibnamefont
  {Dimopoulos}}\ and\ \bibinfo {author} {\bibfnamefont {G.~F.}\ \bibnamefont
  {Giudice}},\ }\href {\doibase 10.1016/0370-2693(96)00390-5} {\bibfield
  {journal} {\bibinfo  {journal} {Physics Letters B}\ }\textbf {\bibinfo
  {volume} {379}},\ \bibinfo {pages} {105} (\bibinfo {year} {1996})},\ \bibinfo
  {note} {arXiv: hep-ph/9602350}\BibitemShut {NoStop}%
\bibitem [{\citenamefont {Antoniadis}\ \emph {et~al.}(1997)\citenamefont
  {Antoniadis}, \citenamefont {Dimopoulos},\ and\ \citenamefont
  {Dvali}}]{antoniadis_millimetre-range_1997}%
  \BibitemOpen
  \bibfield  {author} {\bibinfo {author} {\bibfnamefont {I.}~\bibnamefont
  {Antoniadis}}, \bibinfo {author} {\bibfnamefont {S.}~\bibnamefont
  {Dimopoulos}}, \ and\ \bibinfo {author} {\bibfnamefont {G.}~\bibnamefont
  {Dvali}},\ }\href {\doibase 10.1016/S0550-3213(97)00808-0} {\  (\bibinfo
  {year} {1997}),\ 10.1016/S0550-3213(97)00808-0}\BibitemShut {NoStop}%
\bibitem [{\citenamefont {Feruglio}(2019)}]{Feruglio:2017spp}%
  \BibitemOpen
  \bibfield  {author} {\bibinfo {author} {\bibfnamefont {F.}~\bibnamefont
  {Feruglio}},\ }in\ \href {\doibase 10.1142/9789813238053_0012} {\emph
  {\bibinfo {booktitle} {From My Vast Repertoire ...: Guido Altarelli's
  Legacy}}},\ \bibinfo {editor} {edited by\ \bibinfo {editor} {\bibfnamefont
  {A.}~\bibnamefont {Levy}}, \bibinfo {editor} {\bibfnamefont {S.}~\bibnamefont
  {Forte}}, \ and\ \bibinfo {editor} {\bibfnamefont {G.}~\bibnamefont
  {Ridolfi}}}\ (\bibinfo {year} {2019})\ pp.\ \bibinfo {pages} {227--266},\
  \Eprint {http://arxiv.org/abs/1706.08749} {arXiv:1706.08749 [hep-ph]}
  \BibitemShut {NoStop}%
\bibitem [{\citenamefont {Criado}\ and\ \citenamefont
  {Feruglio}(2018)}]{Criado:2018thu}%
  \BibitemOpen
  \bibfield  {author} {\bibinfo {author} {\bibfnamefont {J.~C.}\ \bibnamefont
  {Criado}}\ and\ \bibinfo {author} {\bibfnamefont {F.}~\bibnamefont
  {Feruglio}},\ }\href@noop {} {\  (\bibinfo {year} {2018})},\ \Eprint
  {http://arxiv.org/abs/1807.01125} {arXiv:1807.01125 [hep-ph]} \BibitemShut
  {NoStop}%
\bibitem [{\citenamefont {Kusenko}\ \emph {et~al.}(2013)\citenamefont
  {Kusenko}, \citenamefont {Loewenstein},\ and\ \citenamefont
  {Yanagida}}]{Kusenko:2012ch}%
  \BibitemOpen
  \bibfield  {author} {\bibinfo {author} {\bibfnamefont {A.}~\bibnamefont
  {Kusenko}}, \bibinfo {author} {\bibfnamefont {M.}~\bibnamefont
  {Loewenstein}}, \ and\ \bibinfo {author} {\bibfnamefont {T.~T.}\ \bibnamefont
  {Yanagida}},\ }\href {\doibase 10.1103/PhysRevD.87.043508} {\bibfield
  {journal} {\bibinfo  {journal} {Phys. Rev.}\ }\textbf {\bibinfo {volume}
  {D87}},\ \bibinfo {pages} {043508} (\bibinfo {year} {2013})},\ \Eprint
  {http://arxiv.org/abs/1209.6403} {arXiv:1209.6403 [hep-ph]} \BibitemShut
  {NoStop}%
\bibitem [{\citenamefont {Acharya}\ \emph {et~al.}(2008)\citenamefont
  {Acharya}, \citenamefont {Kumar}, \citenamefont {Bobkov}, \citenamefont
  {Kane}, \citenamefont {Shao},\ and\ \citenamefont {Watson}}]{Acharya:2008bk}%
  \BibitemOpen
  \bibfield  {author} {\bibinfo {author} {\bibfnamefont {B.~S.}\ \bibnamefont
  {Acharya}}, \bibinfo {author} {\bibfnamefont {P.}~\bibnamefont {Kumar}},
  \bibinfo {author} {\bibfnamefont {K.}~\bibnamefont {Bobkov}}, \bibinfo
  {author} {\bibfnamefont {G.}~\bibnamefont {Kane}}, \bibinfo {author}
  {\bibfnamefont {J.}~\bibnamefont {Shao}}, \ and\ \bibinfo {author}
  {\bibfnamefont {S.}~\bibnamefont {Watson}},\ }\href {\doibase
  10.1088/1126-6708/2008/06/064} {\bibfield  {journal} {\bibinfo  {journal}
  {JHEP}\ }\textbf {\bibinfo {volume} {06}},\ \bibinfo {pages} {064} (\bibinfo
  {year} {2008})},\ \Eprint {http://arxiv.org/abs/0804.0863} {arXiv:0804.0863
  [hep-ph]} \BibitemShut {NoStop}%
\bibitem [{\citenamefont {Acharya}\ \emph {et~al.}(2009)\citenamefont
  {Acharya}, \citenamefont {Kane}, \citenamefont {Watson},\ and\ \citenamefont
  {Kumar}}]{Acharya:2009zt}%
  \BibitemOpen
  \bibfield  {author} {\bibinfo {author} {\bibfnamefont {B.~S.}\ \bibnamefont
  {Acharya}}, \bibinfo {author} {\bibfnamefont {G.}~\bibnamefont {Kane}},
  \bibinfo {author} {\bibfnamefont {S.}~\bibnamefont {Watson}}, \ and\ \bibinfo
  {author} {\bibfnamefont {P.}~\bibnamefont {Kumar}},\ }\href {\doibase
  10.1103/PhysRevD.80.083529} {\bibfield  {journal} {\bibinfo  {journal} {Phys.
  Rev.}\ }\textbf {\bibinfo {volume} {D80}},\ \bibinfo {pages} {083529}
  (\bibinfo {year} {2009})},\ \Eprint {http://arxiv.org/abs/0908.2430}
  {arXiv:0908.2430 [astro-ph.CO]} \BibitemShut {NoStop}%
\bibitem [{\citenamefont {Moroi}\ and\ \citenamefont
  {Randall}(2000)}]{Moroi:1999zb}%
  \BibitemOpen
  \bibfield  {author} {\bibinfo {author} {\bibfnamefont {T.}~\bibnamefont
  {Moroi}}\ and\ \bibinfo {author} {\bibfnamefont {L.}~\bibnamefont
  {Randall}},\ }\href {\doibase 10.1016/S0550-3213(99)00748-8} {\bibfield
  {journal} {\bibinfo  {journal} {Nucl. Phys.}\ }\textbf {\bibinfo {volume}
  {B570}},\ \bibinfo {pages} {455} (\bibinfo {year} {2000})},\ \Eprint
  {http://arxiv.org/abs/hep-ph/9906527} {arXiv:hep-ph/9906527 [hep-ph]}
  \BibitemShut {NoStop}%
\bibitem [{\citenamefont {Allahverdi}\ \emph {et~al.}(2013)\citenamefont
  {Allahverdi}, \citenamefont {Dutta}, \citenamefont {Mohapatra},\ and\
  \citenamefont {Sinha}}]{Allahverdi:2013tca}%
  \BibitemOpen
  \bibfield  {author} {\bibinfo {author} {\bibfnamefont {R.}~\bibnamefont
  {Allahverdi}}, \bibinfo {author} {\bibfnamefont {B.}~\bibnamefont {Dutta}},
  \bibinfo {author} {\bibfnamefont {R.~N.}\ \bibnamefont {Mohapatra}}, \ and\
  \bibinfo {author} {\bibfnamefont {K.}~\bibnamefont {Sinha}},\ }\href
  {\doibase 10.1103/PhysRevLett.111.051302} {\bibfield  {journal} {\bibinfo
  {journal} {Phys. Rev. Lett.}\ }\textbf {\bibinfo {volume} {111}},\ \bibinfo
  {pages} {051302} (\bibinfo {year} {2013})},\ \Eprint
  {http://arxiv.org/abs/1305.0287} {arXiv:1305.0287 [hep-ph]} \BibitemShut
  {NoStop}%
\bibitem [{\citenamefont {Choi}\ \emph {et~al.}(2017)\citenamefont {Choi},
  \citenamefont {Hochberg}, \citenamefont {Kuflik}, \citenamefont {Lee},
  \citenamefont {Mambrini}, \citenamefont {Murayama},\ and\ \citenamefont
  {Pierre}}]{Choi:2017zww}%
  \BibitemOpen
  \bibfield  {author} {\bibinfo {author} {\bibfnamefont {S.-M.}\ \bibnamefont
  {Choi}}, \bibinfo {author} {\bibfnamefont {Y.}~\bibnamefont {Hochberg}},
  \bibinfo {author} {\bibfnamefont {E.}~\bibnamefont {Kuflik}}, \bibinfo
  {author} {\bibfnamefont {H.~M.}\ \bibnamefont {Lee}}, \bibinfo {author}
  {\bibfnamefont {Y.}~\bibnamefont {Mambrini}}, \bibinfo {author}
  {\bibfnamefont {H.}~\bibnamefont {Murayama}}, \ and\ \bibinfo {author}
  {\bibfnamefont {M.}~\bibnamefont {Pierre}},\ }\href {\doibase
  10.1007/JHEP10(2017)162} {\bibfield  {journal} {\bibinfo  {journal} {JHEP}\
  }\textbf {\bibinfo {volume} {10}},\ \bibinfo {pages} {162} (\bibinfo {year}
  {2017})},\ \Eprint {http://arxiv.org/abs/1707.01434} {arXiv:1707.01434
  [hep-ph]} \BibitemShut {NoStop}%
\bibitem [{\citenamefont {Coughlan}\ \emph {et~al.}(1983)\citenamefont
  {Coughlan}, \citenamefont {Fischler}, \citenamefont {Kolb}, \citenamefont
  {Raby},\ and\ \citenamefont {Ross}}]{Coughlan:1983ci}%
  \BibitemOpen
  \bibfield  {author} {\bibinfo {author} {\bibfnamefont {G.~D.}\ \bibnamefont
  {Coughlan}}, \bibinfo {author} {\bibfnamefont {W.}~\bibnamefont {Fischler}},
  \bibinfo {author} {\bibfnamefont {E.~W.}\ \bibnamefont {Kolb}}, \bibinfo
  {author} {\bibfnamefont {S.}~\bibnamefont {Raby}}, \ and\ \bibinfo {author}
  {\bibfnamefont {G.~G.}\ \bibnamefont {Ross}},\ }\href {\doibase
  10.1016/0370-2693(83)91091-2} {\bibfield  {journal} {\bibinfo  {journal}
  {Phys. Lett.}\ }\textbf {\bibinfo {volume} {131B}},\ \bibinfo {pages} {59}
  (\bibinfo {year} {1983})}\BibitemShut {NoStop}%
\bibitem [{\citenamefont {Banks}\ \emph {et~al.}(1994)\citenamefont {Banks},
  \citenamefont {Kaplan},\ and\ \citenamefont {Nelson}}]{Banks:1993en}%
  \BibitemOpen
  \bibfield  {author} {\bibinfo {author} {\bibfnamefont {T.}~\bibnamefont
  {Banks}}, \bibinfo {author} {\bibfnamefont {D.~B.}\ \bibnamefont {Kaplan}}, \
  and\ \bibinfo {author} {\bibfnamefont {A.~E.}\ \bibnamefont {Nelson}},\
  }\href {\doibase 10.1103/PhysRevD.49.779} {\bibfield  {journal} {\bibinfo
  {journal} {Phys. Rev.}\ }\textbf {\bibinfo {volume} {D49}},\ \bibinfo {pages}
  {779} (\bibinfo {year} {1994})},\ \Eprint
  {http://arxiv.org/abs/hep-ph/9308292} {arXiv:hep-ph/9308292 [hep-ph]}
  \BibitemShut {NoStop}%
\bibitem [{\citenamefont {de~Carlos}\ \emph {et~al.}(1993)\citenamefont
  {de~Carlos}, \citenamefont {Casas}, \citenamefont {Quevedo},\ and\
  \citenamefont {Roulet}}]{deCarlos:1993wie}%
  \BibitemOpen
  \bibfield  {author} {\bibinfo {author} {\bibfnamefont {B.}~\bibnamefont
  {de~Carlos}}, \bibinfo {author} {\bibfnamefont {J.~A.}\ \bibnamefont
  {Casas}}, \bibinfo {author} {\bibfnamefont {F.}~\bibnamefont {Quevedo}}, \
  and\ \bibinfo {author} {\bibfnamefont {E.}~\bibnamefont {Roulet}},\ }\href
  {\doibase 10.1016/0370-2693(93)91538-X} {\bibfield  {journal} {\bibinfo
  {journal} {Phys. Lett.}\ }\textbf {\bibinfo {volume} {B318}},\ \bibinfo
  {pages} {447} (\bibinfo {year} {1993})},\ \Eprint
  {http://arxiv.org/abs/hep-ph/9308325} {arXiv:hep-ph/9308325 [hep-ph]}
  \BibitemShut {NoStop}%
\bibitem [{\citenamefont {Binetruy}(2003)}]{Binetruy:2003dx}%
  \BibitemOpen
  \bibfield  {author} {\bibinfo {author} {\bibfnamefont {P.}~\bibnamefont
  {Binetruy}},\ }in\ \href@noop {} {\emph {\bibinfo {booktitle} {{Particle
  physics and cosmology: The interface. Proceedings, NATO Advanced Study
  Institute, School, Cargese, France, August 4-16, 2003}}}}\ (\bibinfo {year}
  {2003})\ pp.\ \bibinfo {pages} {181--234}\BibitemShut {NoStop}%
\bibitem [{\citenamefont {Kawasaki}\ \emph {et~al.}(1999)\citenamefont
  {Kawasaki}, \citenamefont {Kohri},\ and\ \citenamefont
  {Sugiyama}}]{Kawasaki:1999na}%
  \BibitemOpen
  \bibfield  {author} {\bibinfo {author} {\bibfnamefont {M.}~\bibnamefont
  {Kawasaki}}, \bibinfo {author} {\bibfnamefont {K.}~\bibnamefont {Kohri}}, \
  and\ \bibinfo {author} {\bibfnamefont {N.}~\bibnamefont {Sugiyama}},\ }\href
  {\doibase 10.1103/PhysRevLett.82.4168} {\bibfield  {journal} {\bibinfo
  {journal} {Phys. Rev. Lett.}\ }\textbf {\bibinfo {volume} {82}},\ \bibinfo
  {pages} {4168} (\bibinfo {year} {1999})},\ \Eprint
  {http://arxiv.org/abs/astro-ph/9811437} {arXiv:astro-ph/9811437 [astro-ph]}
  \BibitemShut {NoStop}%
\bibitem [{\citenamefont {Kawasaki}\ \emph {et~al.}(2000)\citenamefont
  {Kawasaki}, \citenamefont {Kohri},\ and\ \citenamefont
  {Sugiyama}}]{Kawasaki:2000en}%
  \BibitemOpen
  \bibfield  {author} {\bibinfo {author} {\bibfnamefont {M.}~\bibnamefont
  {Kawasaki}}, \bibinfo {author} {\bibfnamefont {K.}~\bibnamefont {Kohri}}, \
  and\ \bibinfo {author} {\bibfnamefont {N.}~\bibnamefont {Sugiyama}},\ }\href
  {\doibase 10.1103/PhysRevD.62.023506} {\bibfield  {journal} {\bibinfo
  {journal} {Phys. Rev.}\ }\textbf {\bibinfo {volume} {D62}},\ \bibinfo {pages}
  {023506} (\bibinfo {year} {2000})},\ \Eprint
  {http://arxiv.org/abs/astro-ph/0002127} {arXiv:astro-ph/0002127 [astro-ph]}
  \BibitemShut {NoStop}%
\bibitem [{\citenamefont {Hannestad}(2004)}]{Hannestad:2004px}%
  \BibitemOpen
  \bibfield  {author} {\bibinfo {author} {\bibfnamefont {S.}~\bibnamefont
  {Hannestad}},\ }\href {\doibase 10.1103/PhysRevD.70.043506} {\bibfield
  {journal} {\bibinfo  {journal} {Phys. Rev.}\ }\textbf {\bibinfo {volume}
  {D70}},\ \bibinfo {pages} {043506} (\bibinfo {year} {2004})},\ \Eprint
  {http://arxiv.org/abs/astro-ph/0403291} {arXiv:astro-ph/0403291 [astro-ph]}
  \BibitemShut {NoStop}%
\bibitem [{\citenamefont {Giudice}\ \emph {et~al.}(2000)\citenamefont
  {Giudice}, \citenamefont {Kolb},\ and\ \citenamefont
  {Riotto}}]{giudice_largest_2000}%
  \BibitemOpen
  \bibfield  {author} {\bibinfo {author} {\bibfnamefont {G.~F.}\ \bibnamefont
  {Giudice}}, \bibinfo {author} {\bibfnamefont {E.~W.}\ \bibnamefont {Kolb}}, \
  and\ \bibinfo {author} {\bibfnamefont {A.}~\bibnamefont {Riotto}},\ }\href
  {\doibase 10.1103/PhysRevD.64.023508} {\  (\bibinfo {year} {2000}),\
  10.1103/PhysRevD.64.023508}\BibitemShut {NoStop}%
\bibitem [{\citenamefont {Mazumdar}\ and\ \citenamefont
  {Zald{\'\i}var}(2013)}]{mazumdar_quantifying_2014}%
  \BibitemOpen
  \bibfield  {author} {\bibinfo {author} {\bibfnamefont {A.}~\bibnamefont
  {Mazumdar}}\ and\ \bibinfo {author} {\bibfnamefont {B.}~\bibnamefont
  {Zald{\'\i}var}},\ }\href {\doibase 10.1016/j.nuclphysb.2014.07.001} {\
  (\bibinfo {year} {2013}),\ 10.1016/j.nuclphysb.2014.07.001}\BibitemShut
  {NoStop}%
\bibitem [{\citenamefont {Hahn}(2005)}]{hahn_cuba_2005}%
  \BibitemOpen
  \bibfield  {author} {\bibinfo {author} {\bibfnamefont {T.}~\bibnamefont
  {Hahn}},\ }\href {\doibase 10.1016/j.cpc.2005.01.010} {\bibfield  {journal}
  {\bibinfo  {journal} {Computer Physics Communications}\ }\textbf {\bibinfo
  {volume} {168}},\ \bibinfo {pages} {78} (\bibinfo {year} {2005})},\ \bibinfo
  {note} {arXiv: hep-ph/0404043}\BibitemShut {NoStop}%
\bibitem [{\citenamefont {Starobinsky}(1980)}]{starobinsky_new_1980}%
  \BibitemOpen
  \bibfield  {author} {\bibinfo {author} {\bibfnamefont {A.~A.}\ \bibnamefont
  {Starobinsky}},\ }\href {\doibase 10.1016/0370-2693(80)90670-X} {\bibfield
  {journal} {\bibinfo  {journal} {Physics Letters B}\ }\textbf {\bibinfo
  {volume} {91}},\ \bibinfo {pages} {99} (\bibinfo {year} {1980})}\BibitemShut
  {NoStop}%
\bibitem [{\citenamefont {Ellis}\ \emph {et~al.}(2015)\citenamefont {Ellis},
  \citenamefont {Garcia}, \citenamefont {Nanopoulos}, \citenamefont {Olive},\
  and\ \citenamefont {Peloso}}]{ellis_post-inflationary_2015}%
  \BibitemOpen
  \bibfield  {author} {\bibinfo {author} {\bibfnamefont {J.}~\bibnamefont
  {Ellis}}, \bibinfo {author} {\bibfnamefont {M.~A.~G.}\ \bibnamefont
  {Garcia}}, \bibinfo {author} {\bibfnamefont {D.~V.}\ \bibnamefont
  {Nanopoulos}}, \bibinfo {author} {\bibfnamefont {K.~A.}\ \bibnamefont
  {Olive}}, \ and\ \bibinfo {author} {\bibfnamefont {M.}~\bibnamefont
  {Peloso}},\ }\href {\doibase 10.1088/1475-7516/2016/03/008} {\  (\bibinfo
  {year} {2015}),\ 10.1088/1475-7516/2016/03/008}\BibitemShut {NoStop}%
\bibitem [{\citenamefont {Bodeker}(2006)}]{bodeker_moduli_2006}%
  \BibitemOpen
  \bibfield  {author} {\bibinfo {author} {\bibfnamefont {D.}~\bibnamefont
  {Bodeker}},\ }\href {\doibase 10.1088/1475-7516/2006/06/027} {\  (\bibinfo
  {year} {2006}),\ 10.1088/1475-7516/2006/06/027}\BibitemShut {NoStop}%
\end{thebibliography}%

\end{document}